\documentclass[usenatbib,usegraphicx]{mn2e}
\usepackage{amssymb}
\usepackage{morefloats}

\DeclareGraphicsExtensions{.eps,.ps}

\title[Young clusters with a detached pre-main sequence]{Isolating the pre-main sequence in Collinder\,34, NGC\,3293, NGC\,3766 and NGC\,6231}

\author[T.A. Saurin, E. Bica and C. Bonatto]{T.A. Saurin$^{1}$\thanks{E-mail: tiago.saurin@ufrgs.br (TAS)},
E. Bica$^{1}$\thanks{E-mail: bica@if.ufrgs.br (EB)} and C. Bonatto$^{1}$\thanks{E-mail: charles@if.ufrgs.br (CB)}\\
$^{1}$Universidade Federal do Rio Grande do Sul, Departamento de Astronomia, CP\,15051, RS, Porto Alegre 91501-970, Brazil}

\begin{document}

\date{}

\pagerange{\pageref{firstpage}--\pageref{lastpage}} \pubyear{2013}

\maketitle

\label{firstpage}

\begin{abstract}
We employed field star decontaminated 2MASS photometry to study four nearby optical embedded clusters -- Collinder\,34, NGC\,3293, NGC\,3766 and NGC\,6231 -- obtaining deep colour-magnitude diagrams and stellar radial density profiles.
We found what seems to be pre-main sequences detached in different amounts from main sequences in these diagrams.
The structural analysis of each cluster revealed different radial distributions for these two sequences.
We argued that the detached evolutionary sequences in our sample may be evidence of sequential star formation.
Finally, we compared the sample cluster parameters with those of other young clusters in the literature and point out evidence that NGC\,3766 and NGC\,6231 might be evolving to OB associations.
\end{abstract}
\begin{keywords}
open clusters and associations: individual: Collinder\,34 --
open clusters and associations: individual: NGC\,3293 --
open clusters and associations: individual: NGC\,3766 --
open clusters and associations: individual: NGC\,6231
\end{keywords}

\section{Introduction}
\label{sec:in}

Young star clusters remain embedded in molecular clouds during the first
$\sim$10\,Myr of their existence, so that their stellar content should be
observed in infrared wavelengths \citep{lada03}. During this stage, they
have a lot of stars that do not fuse hydrogen yet, the pre-main sequence
(PMS) stars distinguished from the main-sequence (MS) stars.

As plotted in a colour-magnitude diagram (CMD), these two classes of stars
are distributed along different tracks that can be modeled by means of
theoretical isochrones (e.g. \citealt*{siess00}; \citealt{marigo08}). Stars
with colour and magnitudes occurring along the same curve are expected
to be coeval. Factors such as binarity, variability, differential reddening,
photometric uncertainties and continuous star formation cause spread
around these isochrones (\citealt*{bbl12a}; \citealt*{bbl12b}).
Photometry of PMS stars is specially affected by these factors. Among them
stand out the T Tauri, young variables with circumstellar
disks and infrared excess emission \citep{furlan09}.

Snapshots of the dynamical stages of a star cluster may be obtained from
radial density profiles (RDPs). Embedded clusters are not expected to have
profiles following an isothermal sphere model (e.g. \citealt{kin62}).
Nevertheless, some of these density profiles are fitted by a King-like
model providing reliable structural parameters. RDP analyses can also
reveal clues on cluster dissolution when extended profiles are observed
(e.g. \citealt*{sbb12}) and gradients in the ratios of high-to-low-mass and
old-to-young stars that may be interpreted as mass and age segregation
(e.g. \citealt{hill97}).

Using Two Micron All Sky Survey
(2MASS\footnote{\it http://www.ipac.caltech.edu/2mass}; \citealt{skruts06})
photometry we analyze the evolutionary sequences and radial profiles of a
sample of young clusters in order to investigate triggered/sequential star
formation (e.g. \citealt*{sbb10}).
The sample consists of objects catalogued by \citet{col31}
who has established some of the bases of the modern cluster analysis.
We selected four clusters, the little studied Collinder\,34, plus
NGC\,3293 (Collinder\,224), NGC\,3766 (Collinder\,248) and
NGC\,6231 (Collinder\,315) that are bright nearby young clusters
previously studied in various wavelengths (e.g. \citealt{dias02};
\citealt{karc13}).
In the present work we perform a field star decontamination procedure
in order to isolate and go deeper than previous studies into the PMS
of these clusters.

This paper is organized as follows. In Sect.\,\ref{sec:dec} we explain the
field star decontamination method. In Sect.\,\ref{sec:cmd} we build
CMDs and estimate astrophysical parameters. In
Sect.\,\ref{sec:rdp} we analyze the radial density profiles and derive
structural parameters. In Sect.\,\ref{sec:dis} we discuss the results. Finally,
in Sect.\,\ref{sec:con} we give the concluding remarks.

\begin{table}
\caption[]{Adopted central coordinates for the clusters. By columns:
(1) star cluster identification;
(2) Galactic longitude;
(3) Galactic latitude;
(4) right ascension (J2000);
(5) declination (J2000).}
\label{tab:coo}
\renewcommand{\tabcolsep}{1.8mm}
\renewcommand{\arraystretch}{1.25}
\centering
\begin{tabular}{lcccc}
\hline
\hline
\multicolumn{1}{c}{cluster}&\multicolumn{1}{c}{$l$} &\multicolumn{1}{c}{$b$}&\multicolumn{1}{c}{$\alpha$} &\multicolumn{1}{c}{$\delta$}\\
\multicolumn{1}{c}{(1)} &\multicolumn{1}{c}{(2)}&\multicolumn{1}{c}{(3)} &\multicolumn{1}{c}{(4)} &\multicolumn{1}{c}{(5)}\\
\hline
Collinder\,34 & 138\fdg03 & +1\fdg50 & 02$^{\rmn{h}}$59$^{\rmn{m}}$23\fs17 & +60\degr33\arcmin59\farcs5\\
NGC\,3293 & 285\fdg86 & +0\fdg07 & 10$^{\rmn{h}}$35$^{\rmn{m}}$53\fs10 & $-$58\degr13\arcmin55\farcs5\\
NGC\,3766 & 294\fdg12 & $-$0\fdg03 & 11$^{\rmn{h}}$36$^{\rmn{m}}$14\fs00 & $-$61\degr36\arcmin30\farcs0\\
NGC\,6231 & 343\fdg46 & +1\fdg18 & 16$^{\rmn{h}}$54$^{\rmn{m}}$10\fs06 & $-$41\degr49\arcmin30\farcs1\\
\hline
\end{tabular}
\end{table}

\section{2MASS photometry and field star decontamination}
\label{sec:dec}

We used the catalogue II/246 \citep{cutri03} available in the
VizieR\footnote{\it http://vizier.u-strasbg.fr/viz-bin/VizieR} data base
\citep*{ocbama00} to extract the 2MASS photometry of the clusters in the
$J$, $H$ and $K_S$ bands within large circular areas ({\it r}\,=\,100\,arcmin).
We restricted our analyses to stars with photometric errors $\leq$\,0.1\,mag
in each band.

The cluster coordinates in WEBDA\footnote{\it http://webda.physics.muni.cz} were
taken as the starting point to find an optimized centre, which corresponds to
the RDP with the highest stellar density at the innermost (radial) bin
(Table\,\ref{tab:coo}).
We found that Collinder\,34 was better centred on its brightest star,
HD\,18326; for NGC\,3293, we chose the centre near the Be star V439\,Car; for
NGC\,3766, the WEBDA coordinates were kept. Finally, for NGC\,6231, the
coordinates of the eclipsing binary star V1007\,Sco proved to be adequate.

Field star contamination is expected to be present in essentially all cluster
observations, especially when the object is close to the Galactic plane.
Therefore, it is necessary to isolate the probable cluster members. We use
a method that performs a comparison between the photometric properties of the
stars in an offset field and the cluster area
(e.g. \citealt{bonbic07}; \citealt{bonbic10}). It works according to the
following algorithm:

\begin{enumerate}\itemsep5pt
\item A circular area of radius 10\,arcmin (or 20\,arcmin) of the cluster and an
annular area (80\,$\leq$\,{\it r}\,$\leq$\,100\,arcmin) of the field were
selected. The contamination in the cluster area was modeled by comparison of
three-dimensional CMDs $J$$\times${\it (J$-$H)}$\times${\it (J$-$K$_S$)} of
both regions. These CMDs were divided by cells of dimensions
${\it \Delta} J$\,=\,1.0 and
${\it \Delta}(J$$-$$H)$\,=\,${\it \Delta} (J$$-$$K_S)$\,=\,0.2, where
the probability of a star to be member was computed. Thus, the density of
stars was obtained in each cell of the cluster and field CMDs.

\item The comparison field density was converted into an
integer number of stars to be subtracted from that in the
cluster area, on a cell-by-cell basis, resulting the number of member stars in
each cell ($N^{cell}_{clean}$). Shifts in the cell positioning corresponding to
1/3 of the adopted cell size in each dimension were allowed, so that 243
different setups were run, each one with a total number of member stars
$N_{mem}\,=\,\sum_{cell}N^{cell}_{clean}$. By averaging these values
the expected number of member stars was computed within the extraction area of
the cluster ($\left<N_{mem}\right>$).

\item The $\left<N_{mem}\right>$ stars that have more often survived the 243
runs were considered cluster members and transposed to the decontaminated CMD.
The difference between the expected number of field stars (sometimes
fractional) and the number of stars effectively subtracted (integer) from each
cell provides the subtraction efficiency.
\end{enumerate}

The resulting subtraction efficiencies were
90.4$\pm$1.8 per cent for Collinder\,34,
83.4$\pm$1.0 per cent for NGC\,3293,
92.5$\pm$0.5 per cent for NGC\,3766,
and 82.8$\pm$0.9 per cent for NGC\,6231.

We emphasize that the clean CMDs are essential guides for the
optimal setting of theoretical isochrone curves and estimating
reliable parameters (Sect.\,\ref{sec:cmd}). Circular areas of
different radii were tested, but we found that larger
radii resulted in lower subtraction efficiencies, while smaller
radii removed too many points from the CMDs -- including stars catalogued
as members.

\section{Colour-magnitude diagrams}
\label{sec:cmd}

\begin{figure*}
\begin{minipage}{0.685\textheight}
\resizebox{\hsize}{!}{\includegraphics{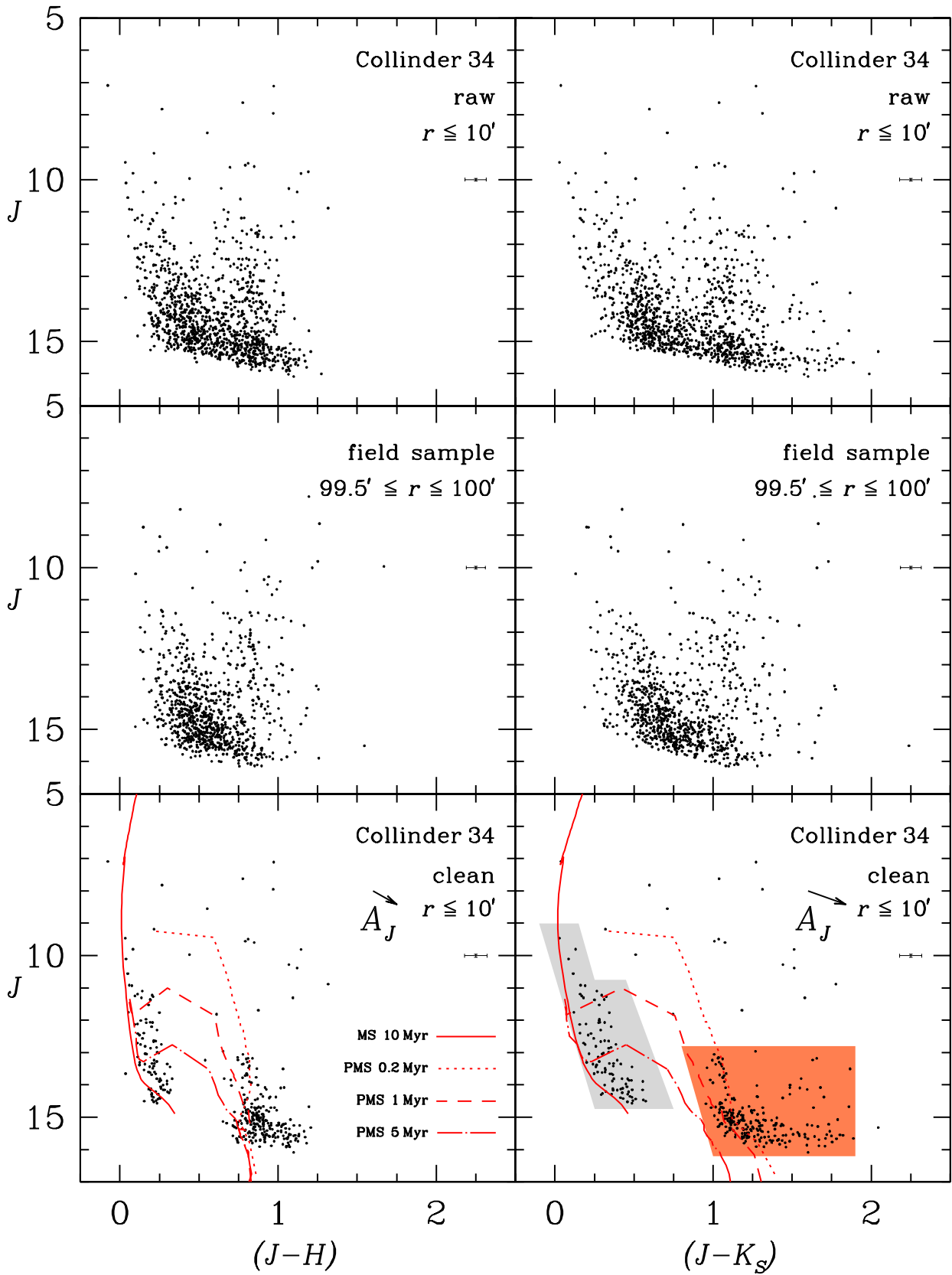}}
\caption[]{Collinder\,34: 2MASS colour-magnitude diagrams extracted inside {\it r}\,$\leq$\,10\,arcmin.
Mean uncertainties are represented by error bars.
Top panels: observed photometry with {\it J$\times$(J$-$H)} (left) and {\it J$\times$(J$-$K$_S$)} (right).
Middle: field equal-area extraction.
Bottom: decontaminated CMDs with 10\,Myr Solar-metallicity Padova isochrone and 0.2, 1 and 5\,Myr PMS isochrones.
Light-shaded polygon: MS colour-magnitude filter.
Heavy-shaded polygon: the same for the PMS stars.
The arrows indicate the reddening vectors for $A_J$\,=\,0.42\,mag.}
\label{fig:cr34_cmd}
\end{minipage}
\end{figure*}

\begin{figure*}
\begin{minipage}{0.685\textheight}
\resizebox{\hsize}{!}{\includegraphics{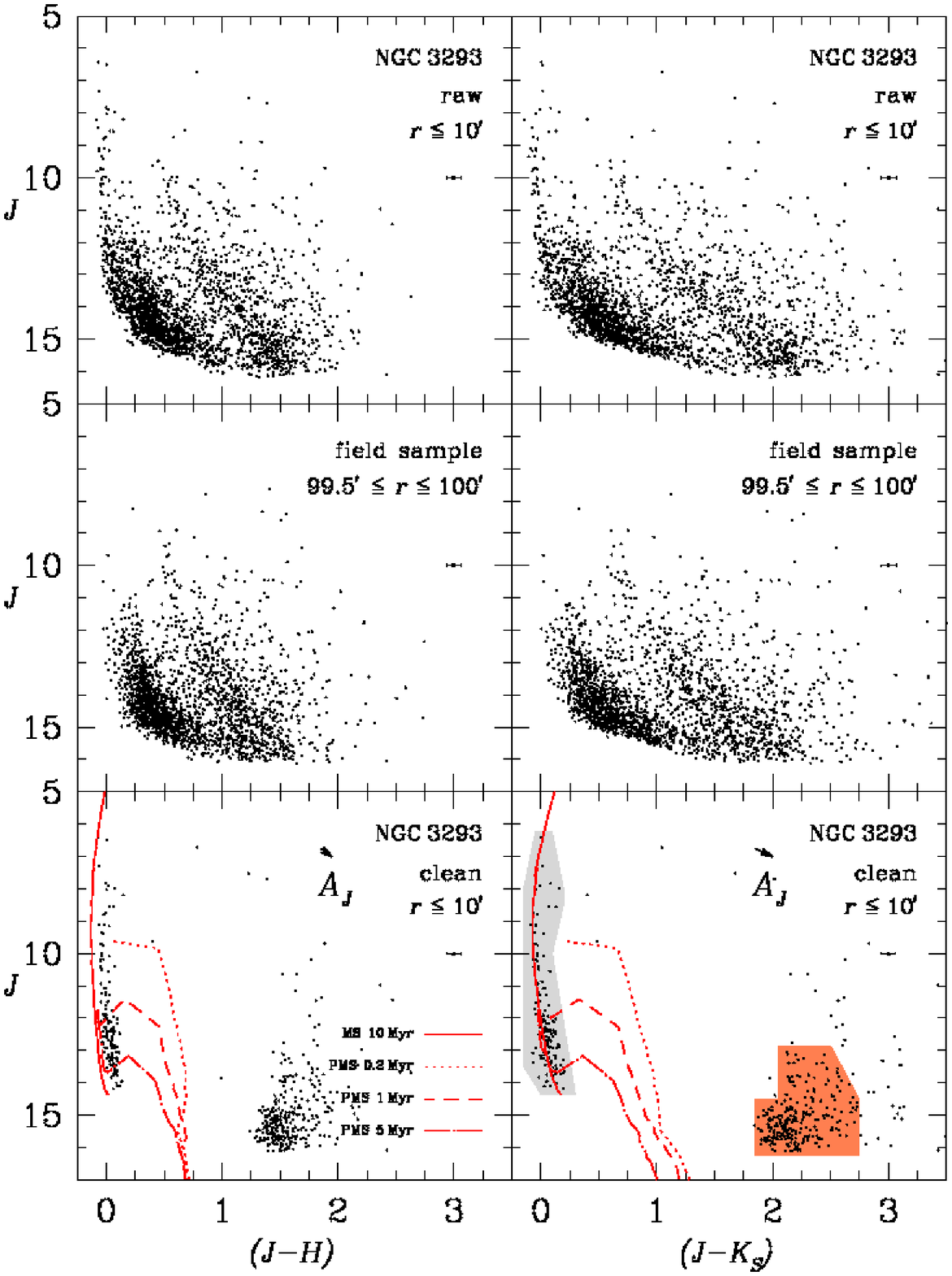}}
\caption[]{NGC\,3293: similar to Fig.\,\ref{fig:cr34_cmd}.
The arrows indicate the reddening vectors for $A_J$\,=\,0.26\,mag.}
\label{fig:ngc3293_cmd}
\end{minipage}
\end{figure*}

\begin{figure*}
\begin{minipage}{0.685\textheight}
\resizebox{\hsize}{!}{\includegraphics{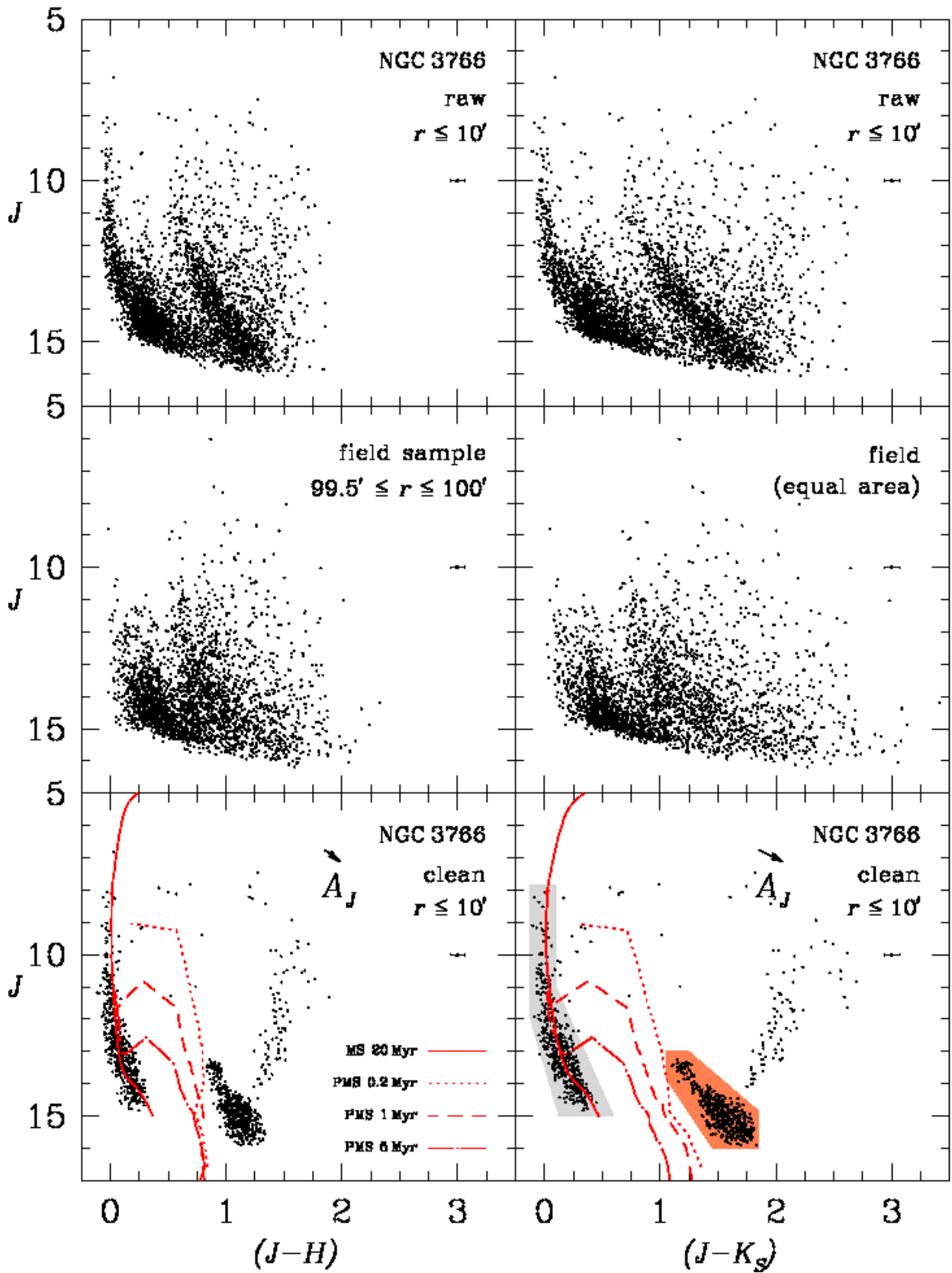}}
\caption[]{NGC\,3766: similar to Fig.\,\ref{fig:cr34_cmd},
but with a 20\,Myr Padova isochrone.
The arrows indicate the reddening vectors for $A_J$\,=\,0.35\,mag.}
\label{fig:ngc3766_cmd}
\end{minipage}
\end{figure*}

\begin{figure*}
\begin{minipage}{0.685\textheight}
\resizebox{\hsize}{!}{\includegraphics{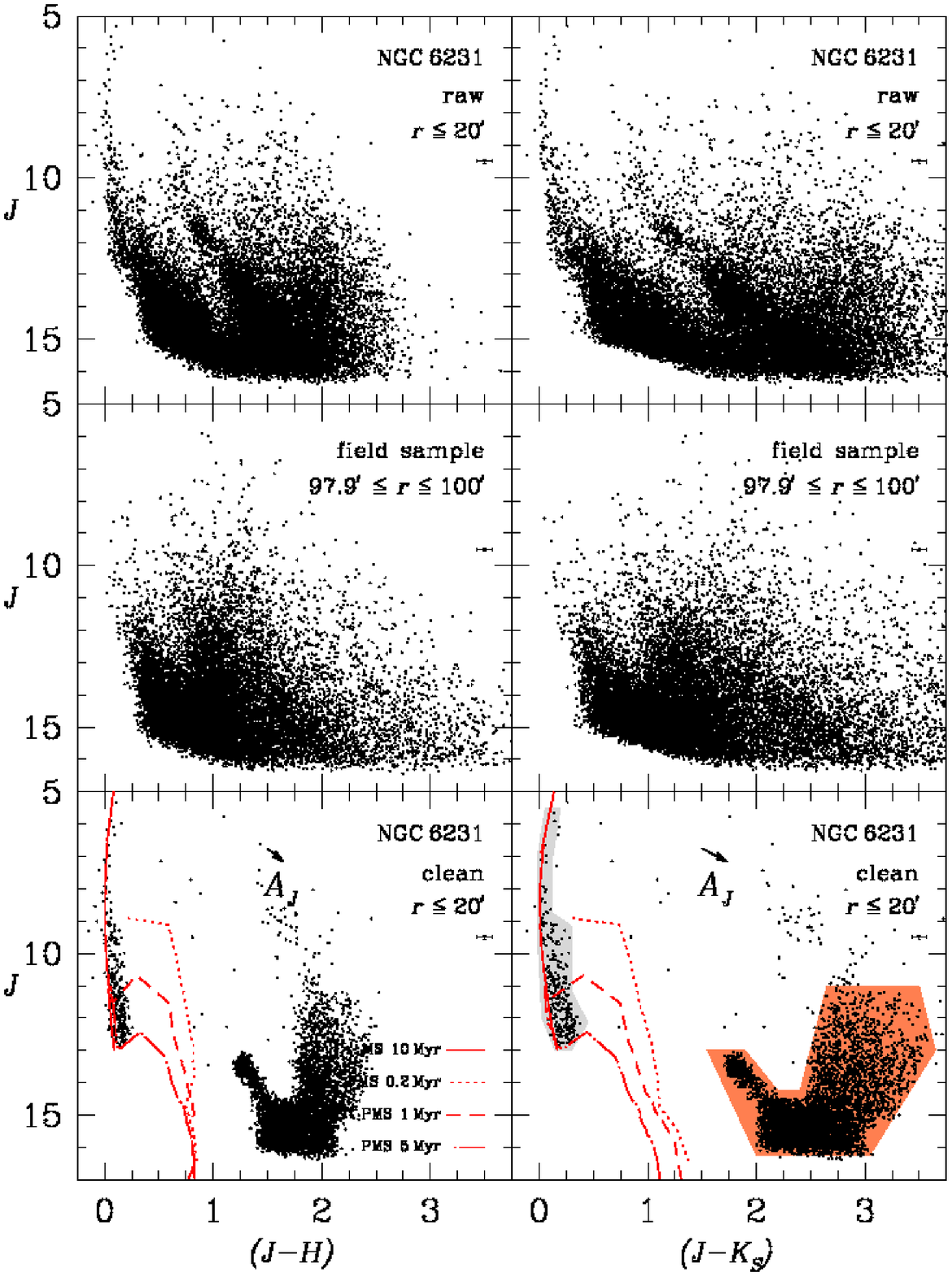}}
\caption[]{NGC\,6231: similar to Fig.\,\ref{fig:cr34_cmd}.
The arrows indicate the reddening vectors for $A_J$\,=\,0.40\,mag.}
\label{fig:ngc6231_cmd}
\end{minipage}
\end{figure*}

In Figs.\,\ref{fig:cr34_cmd} to \ref{fig:ngc6231_cmd} are shown the 2MASS
colour-magnitude diagrams {\it J$\times$(J$-$H)} and {\it J$\times$(J$-$K$_S$)}
of the four clusters. The CMDs present raw cluster photometry within the
circular areas (Sect.\,\ref{sec:dec}),
field star photometry within an annulus of equal area,
and cluster decontaminated photometry.

The decontamination procedure enhances the PMS and enables to detect the low
mass limit of the MS, as expected from PMS to MS evolution. This cutoff is
detected for all sample clusters
(Figs.\,\ref{fig:cr34_cmd}-\ref{fig:ngc6231_cmd}).
An important feature in the CMDs of the sample is the
colour separation between the MS and probable PMS. This feature
has also been observed in Trumpler\,37 \citep{sbb12}.

2MASS photometric uncertainties and degeneracies do not allow to detect
metallicity differences for young populations. Thus, we adopted
Solar-metallicity isochrones, suitable for the Galactic disc in
general. For the MS of all clusters we used 10\,Myr or 20\,Myr Padova
isochrones \citep{marigo08} plus 0.2, 1 and 5\,Myr PMS isochrones
of \citet{siess00} converted to 2MASS photometry \citep{kenhar95}.

For each cluster, reddening and distance have been obtained by means of a
spectral type calibration using a catalogued
blue star, avoiding
spectroscopic binaries and variables (Table\,\ref{tab:stars}).
The following stars listed in the
SIMBAD\footnote{\it http://simbad.u-strasbg.fr/simbad} astronomical database
\citep{wen00} have been used in spectral type calibrations:
TYC\,4048-1432-1 for Collinder\,34, CPD-57\,3520 for NGC\,3293, CPD-60\,3120
for NGC\,3766, and SBL\,164 for NGC\,6231. Table\,\ref{tab:cmd}
shows the parameters {\it (m$-$M)J}, {\it E(J$-$H)} and {\it E(J$-$K$_S$)},
obtained from the MS isochrone solutions, and {\it E(H$-$K$_S$)}, $A_J$,
{\it E(B$-$V)} and $A_V$, estimated using the relations given by
\citet*{dusabi02}.

Heliocentric and Galactocentric distances for each cluster have been estimated
using the photometric parameters and assuming that the distance from the Sun to
the Galactic centre is {\it R$_{\sun}$}\,=\,8.4$\pm$0.6\,kpc
\citep{reid09}. The resulting values are in Table\,\ref{tab:par}.

A mass estimate of the MS stellar content was made by summing the
individual masses of each star according to the
mass-luminosity relation for dereddened colours and magnitudes.
Collinder\,34 has MS stellar masses in the range 1.4\,M$_{\sun}$ to
17.8\,M$_{\sun}$, and its MS mass amounts to
359$^{+37}_{-27}$\,M$_{\sun}$.
For NGC\,3293, the range is 1.7\,M$_{\sun}$ to 17.8\,M$_{\sun}$,
yielding a MS mass of 899$^{+11}_{-8}$\,M$_{\sun}$.
For NGC\,3766, the range is 1.2\,M$_{\sun}$ to 11.2\,M$_{\sun}$,
yielding a MS mass of 1682$\pm$31\,M$_{\sun}$.
For NGC\,6231, the range is 2.5\,M$_{\sun}$ to 17.8\,M$_{\sun}$,
yielding a MS mass of 1939$^{+17}_{-13}$\,M$_{\sun}$.
A similar procedure is not possible for the PMS, so we simply counted the number
of stars and multiplied it by a mean PMS stellar mass. This is 0.6\,M$_{\sun}$
by assuming an initial mass function of \citet{kroupa01} in the range
0.08\,M$_{\sun}$ to 7\,M$_{\sun}$.
All these masses were derived within circular areas of radius 10 or 20\,arcmin
(Sect.\,\ref{sec:dec}) and are summarized in Table\,\ref{tab:par}.
Note that these values must be lower limits, owing to
the presence of dust and gas, unresolved binaries, and the fact that we have
considered only the more probable cluster members (decontaminated CMDs).

\begin{figure*}
\begin{minipage}[t]{0.45\linewidth}
\centering
\resizebox{\hsize}{!}{\includegraphics{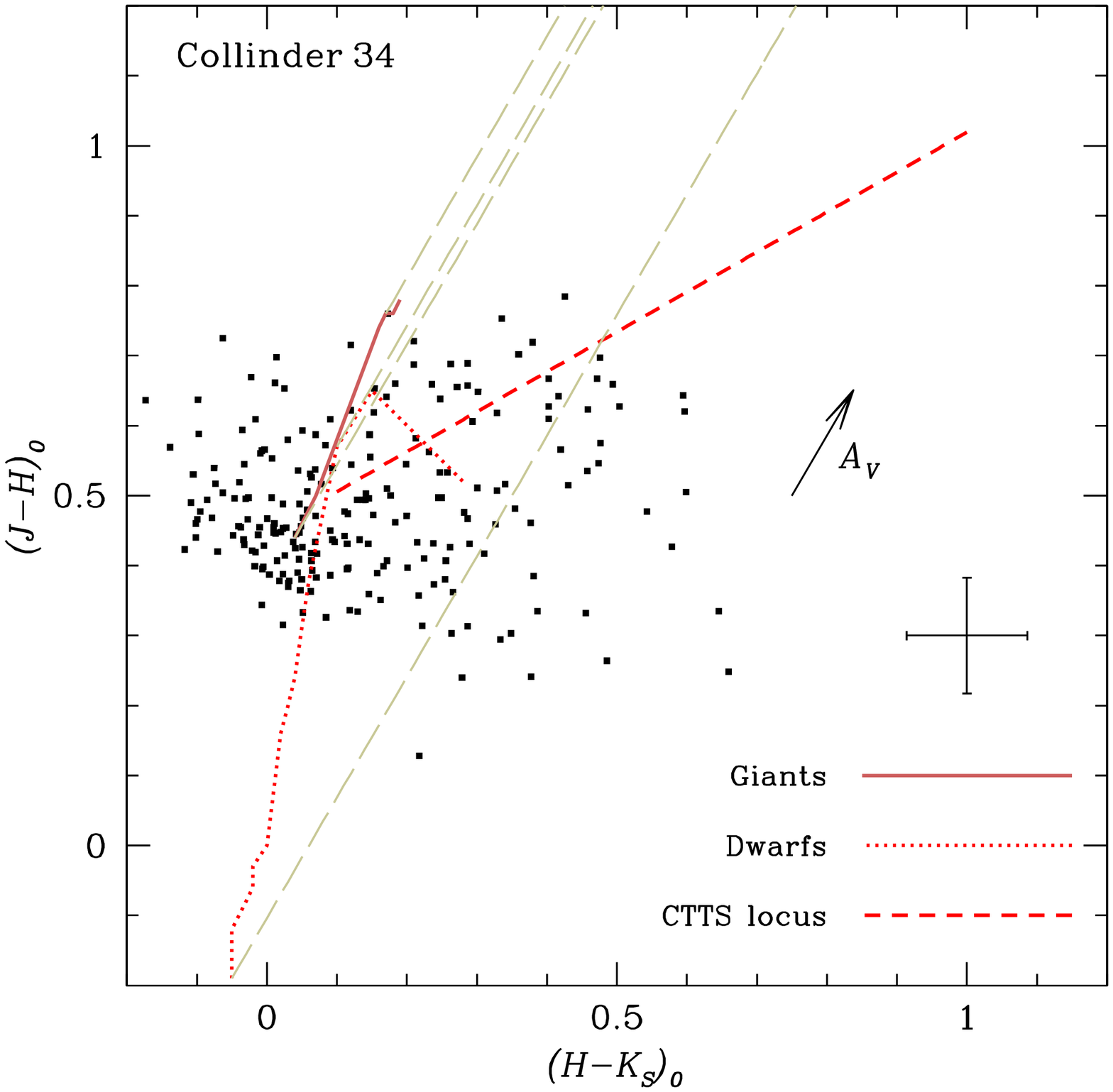}}
\caption[]{Collinder\,34: 2MASS extinction-corrected 2CD
of the probable PMS stars. Mean uncertainties are represented by the error bars.
Solid line is the sequence of giants, dotted line is the sequence
of dwarfs, and short-dashed line is the classical T Tauri stars locus. The
arrow indicates the reddening vector for $A_V$\,=\,1.51\,mag. The
long-dashed lines encompass a strip area parallel to the reddening vector. The stars with
colours outside and to the right of the strip area have infrared excess
emission.}
\label{fig:cr34_2cd}
\end{minipage}
\hspace{0.15cm}
\begin{minipage}[t]{0.45\linewidth}
\centering
\resizebox{\hsize}{!}{\includegraphics{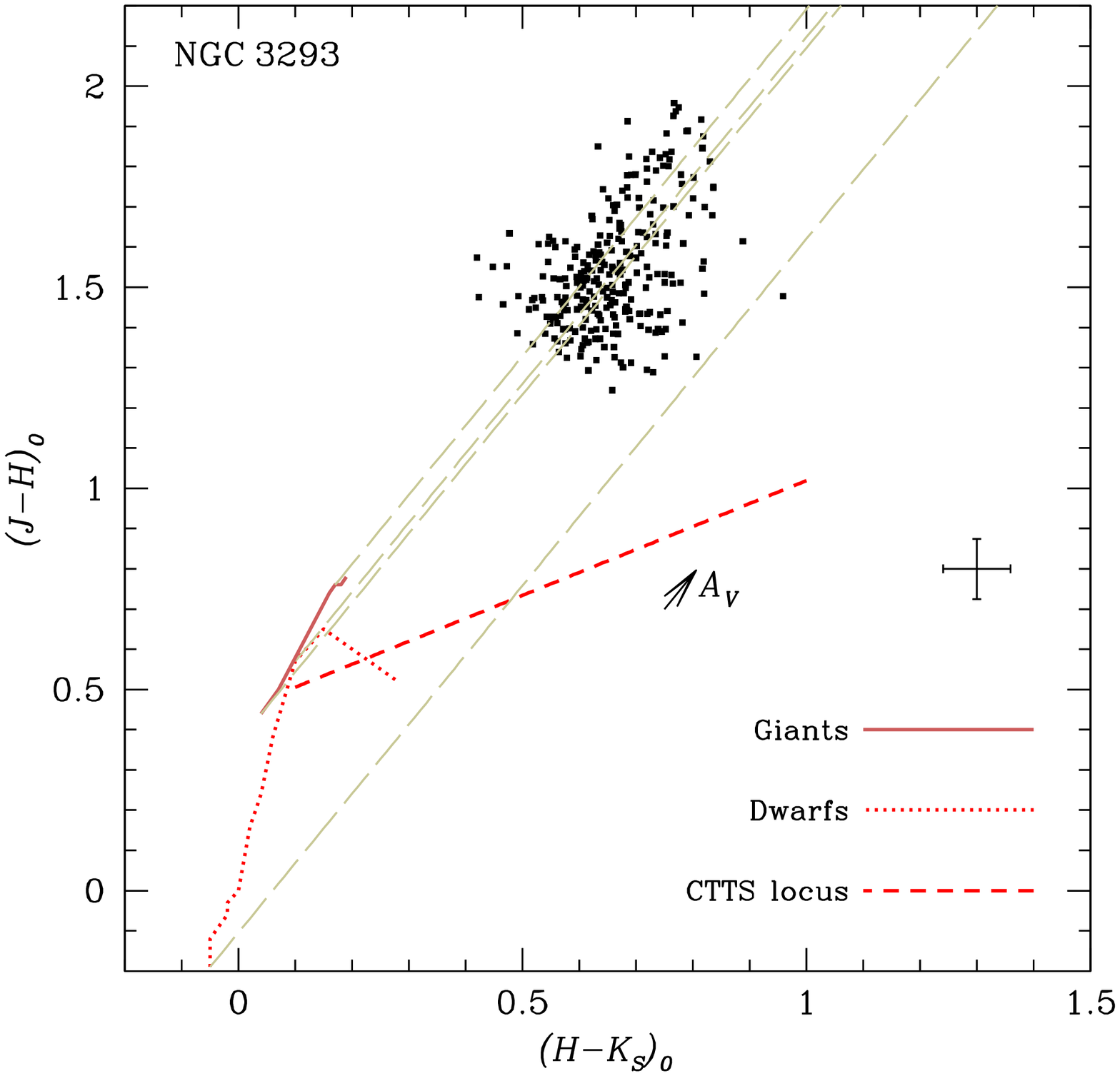}}
\caption[]{NGC\,3293: same as Fig.\,\ref{fig:cr34_2cd}.
The arrow indicates the reddening vector for $A_V$\,=\,0.96\,mag.}
\label{fig:ngc3293_2cd}
\end{minipage}
\begin{minipage}[t]{0.45\linewidth}
\centering
\resizebox{\hsize}{!}{\includegraphics{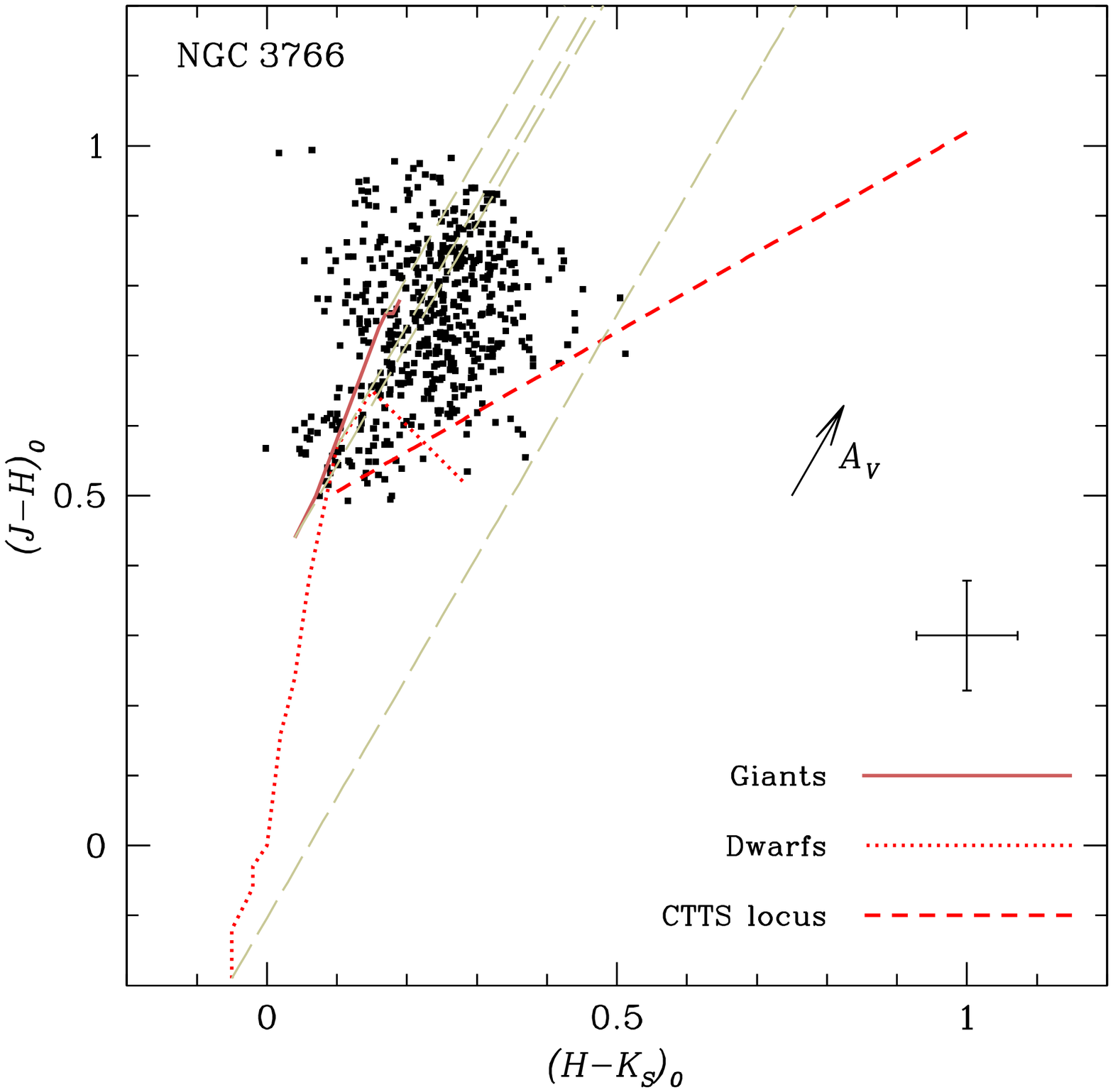}}
\caption[]{NGC\,3766: same as Fig.\,\ref{fig:cr34_2cd}.
The arrow indicates the reddening vector for $A_V$\,=\,1.28\,mag.}
\label{fig:ngc3766_2cd}
\end{minipage}
\hspace{0.15cm}
\begin{minipage}[t]{0.45\linewidth}
\centering
\resizebox{\hsize}{!}{\includegraphics{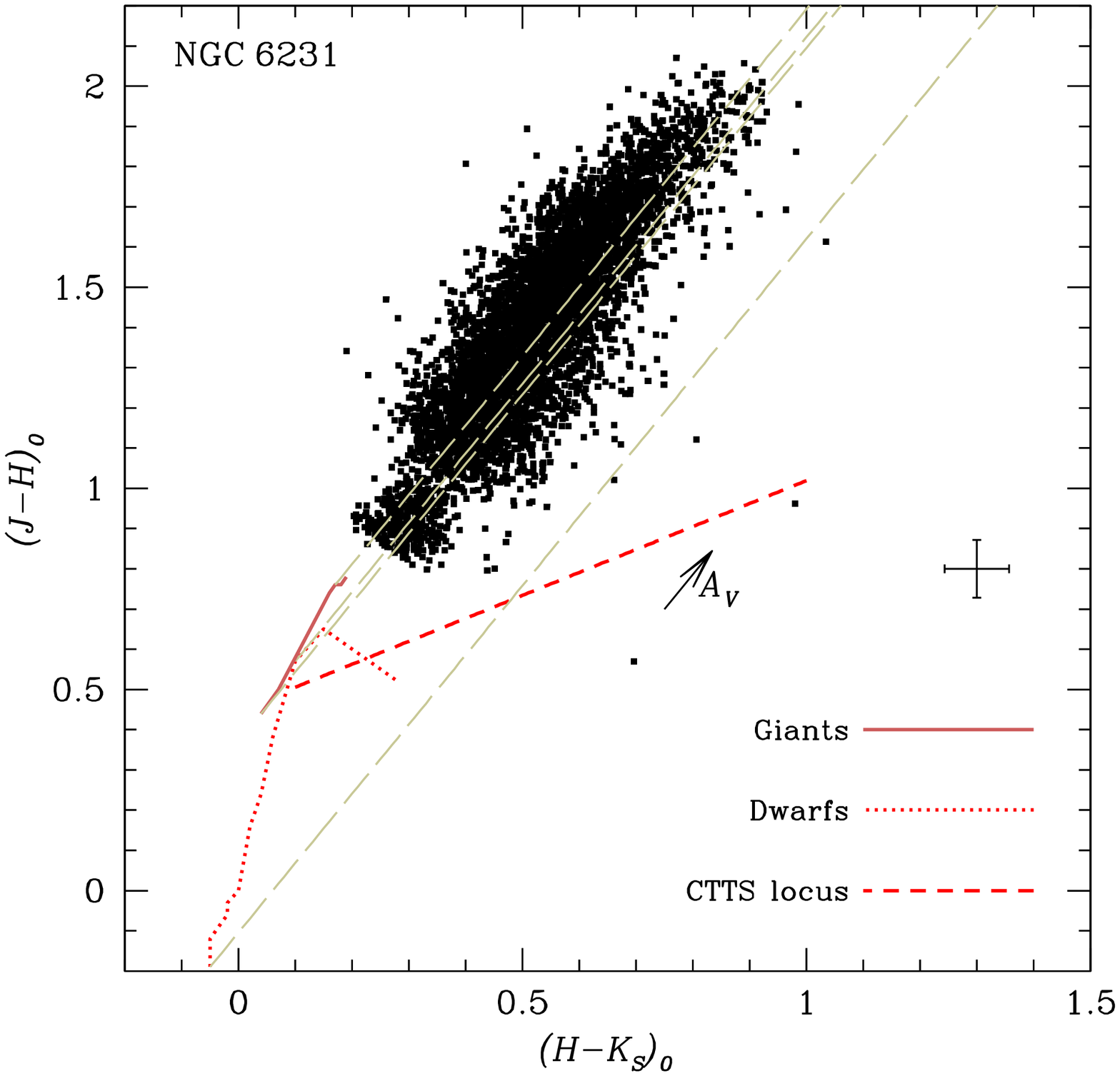}}
\caption[]{NGC6231: same as Fig.\,\ref{fig:cr34_2cd}.
The arrow indicates the reddening vector for $A_V$\,=\,1.44\,mag.}
\label{fig:ngc6231_2cd}
\end{minipage}
\end{figure*}

Additional comparisons with sequences of dwarfs and giants \citep{sk82},
and the classical T Tauri stars (CTTSs) locus \citep{meye97} -- transformed
into the 2MASS photometric system using the relations derived by \citet{car01}
-- in extinction corrected colour-colour diagrams (2CDs)
show a significant population in the PMS locus
(Figs.\,\ref{fig:cr34_2cd}-\ref{fig:ngc6231_2cd}).

\begin{table*}
\begin{minipage}{1\textwidth}
\caption[]{Reference stars used for the spectral type calibration of the theoretical isochrones. By columns:
(1) star identification;
(2) associated cluster;
(3) right ascension (J2000);
(4) declination (J2000);
(5) {\it J}-band apparent magnitude;
(6) {\it J$-$H} colour;
(7) {\it J$-$K$_S$} colour;
(8) spectral type.
}
\label{tab:stars}
\renewcommand{\tabcolsep}{1.8mm}
\renewcommand{\arraystretch}{1.25}
\centering
\begin{tabular}{lcccrrrc}
\hline
\hline
\multicolumn{1}{c}{star} &\multicolumn{1}{c}{cluster}
&\multicolumn{1}{c}{$\alpha$} &\multicolumn{1}{c}{$\delta$}
&\multicolumn{1}{c}{\it m$_J$} &\multicolumn{1}{c}{\it J$-$H} &\multicolumn{1}{c}{\it J$-$K$_S$} &\multicolumn{1}{c}{ST}\\
\multicolumn{1}{c}{(1)} &\multicolumn{1}{c}{(2)} &\multicolumn{1}{c}{(3)}
&\multicolumn{1}{c}{(4)} &\multicolumn{1}{c}{(5)} &\multicolumn{1}{c}{(6)} &\multicolumn{1}{c}{(7)} &\multicolumn{1}{c}{(8)}\\
\hline
TYC\,4048-1432-1 & Collinder\,34 & 02$^{\rmn{h}}$59$^{\rmn{m}}$21\fs19 &   +60\degr25\arcmin22\farcs0 &  9.464$\pm$0.022 &    0.034$\pm$0.036 &    0.029$\pm$0.030 & B8\\
CPD-57\,3520     & NGC\,3293     & 10$^{\rmn{h}}$35$^{\rmn{m}}$56\fs61 & $-$58\degr12\arcmin40\farcs9 & 10.197$\pm$0.027 & $-$0.022$\pm$0.037 & $-$0.061$\pm$0.035 & B2\\
CPD-60\,3120     & NGC\,3766     & 11$^{\rmn{h}}$36$^{\rmn{m}}$08\fs25 & $-$61\degr34\arcmin19\farcs4 & 10.661$\pm$0.026 & 0.005$\pm$0.037 & 0.031$\pm$0.033 & B4\\
SBL\,164         & NGC\,6231     & 16$^{\rmn{h}}$53$^{\rmn{m}}$46\fs99 & $-$41\degr48\arcmin55\farcs0 & 11.683$\pm$0.023 & 0.058$\pm$0.038 & 0.036$\pm$0.045 & B7\\
\hline
\end{tabular}
\end{minipage}
\end{table*}

\begin{table*}
\begin{minipage}{1\textwidth}
\caption[]{Photometric parameters of each cluster. By columns:
(1) star cluster identification;
(2) distance modulus;
(3),(4),(5) 2MASS colour excesses;
(6) {\it J} band absorption;
(7) {\it E(B$-$V)} colour excess;
(8) {\it V} band absorption.}
\label{tab:cmd}
\renewcommand{\tabcolsep}{1.8mm}
\renewcommand{\arraystretch}{1.25}
\centering
\begin{tabular}{lccccccc}
\hline
\hline
\multicolumn{1}{c}{cluster} &\multicolumn{1}{c}{\it (m$-$M)J} &\multicolumn{1}{c}{\it E(J$-$H)}
&\multicolumn{1}{c}{\it E(J$-$K$_S$)} &\multicolumn{1}{c}{\it E(H$-$K$_S$)} &\multicolumn{1}{c}{\it A$_J$}
&\multicolumn{1}{c}{\it E(B$-$V)} &\multicolumn{1}{c}{\it A$_V$}\\
\multicolumn{1}{c}{(1)} &\multicolumn{1}{c}{(2)} &\multicolumn{1}{c}{(3)}
&\multicolumn{1}{c}{(4)} &\multicolumn{1}{c}{(5)} &\multicolumn{1}{c}{(6)} &\multicolumn{1}{c}{(7)} &\multicolumn{1}{c}{(8)}\\
\hline
Collinder\,34 & 12.03$\pm$0.19 & 0.15$\pm$0.03 & 0.24$\pm$0.05 & 0.09$\pm$0.02 & 0.42$\pm$0.08 & 0.49$\pm$0.10 & 1.51$\pm$0.30\\
NGC\,3293 & 12.44$\pm$0.03 & 0.10$\pm$0.04 & 0.15$\pm$0.06 & 0.06$\pm$0.02 & 0.26$\pm$0.10 & 0.31$\pm$0.12 & 0.96$\pm$0.37\\
NGC\,3766 & 11.85$\pm$0.03 & 0.13$\pm$0.05 & 0.20$\pm$0.08 & 0.07$\pm$0.03 & 0.35$\pm$0.14 & 0.41$\pm$0.16 & 1.28$\pm$0.50\\
NGC\,6231 & 11.71$\pm$0.02 & 0.14$\pm$0.04 & 0.23$\pm$0.06 & 0.08$\pm$0.02 & 0.40$\pm$0.11 & 0.46$\pm$0.13 & 1.44$\pm$0.40\\
\hline
\end{tabular}
\end{minipage}
\end{table*}

\section{Radial density profiles}
\label{sec:rdp}

In order to build a stellar RDP of each cluster over its whole extension
(beyond 10 or 20\,arcmin) with minimum
contamination, we defined colour-magnitude filters (shaded areas in the
decontaminated CMDs of Figs.\,\ref{fig:cr34_cmd}-\ref{fig:ngc6231_cmd}). They
were applied to the raw photometry within circular areas of
radius 100\,arcmin. Because the sample clusters show detached evolutionary sequences
in the CMDs, we defined separate filters for each one.

\begin{figure*}
\begin{minipage}{0.685\textheight}
\resizebox{\hsize}{!}{\includegraphics{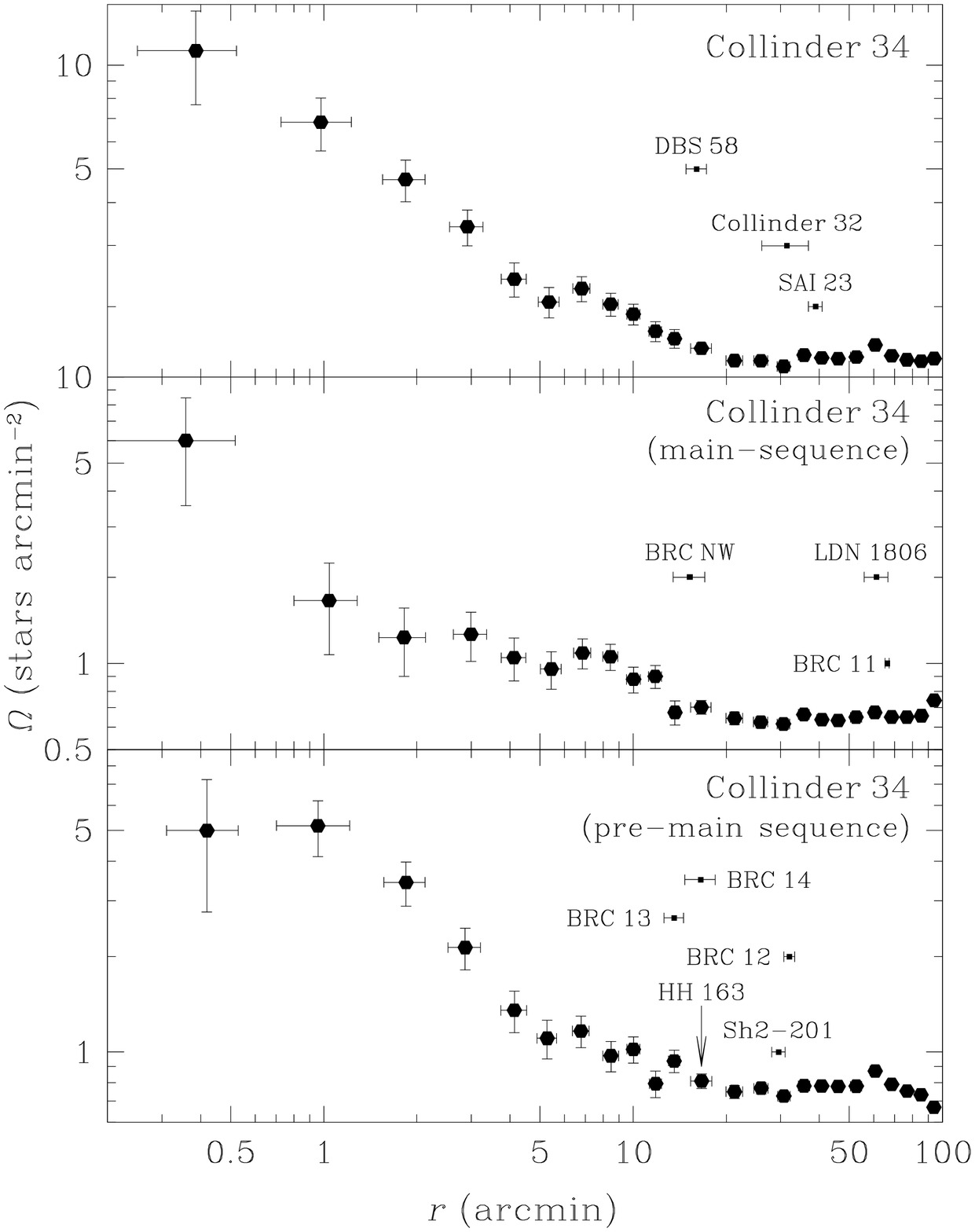}}
\caption[]{Collinder\,34: stellar RDPs built with stars
filtered by colour and magnitude (Fig.\,\ref{fig:cr34_cmd}).
Top panel: total profile (MS\,+\,PMS).
Middle panel: MS profile.
Bottom panel: PMS profile.}
\label{fig:cr34_rdp}
\end{minipage}
\end{figure*}

\begin{figure*}
\begin{minipage}{0.685\textheight}
\resizebox{\hsize}{!}{\includegraphics{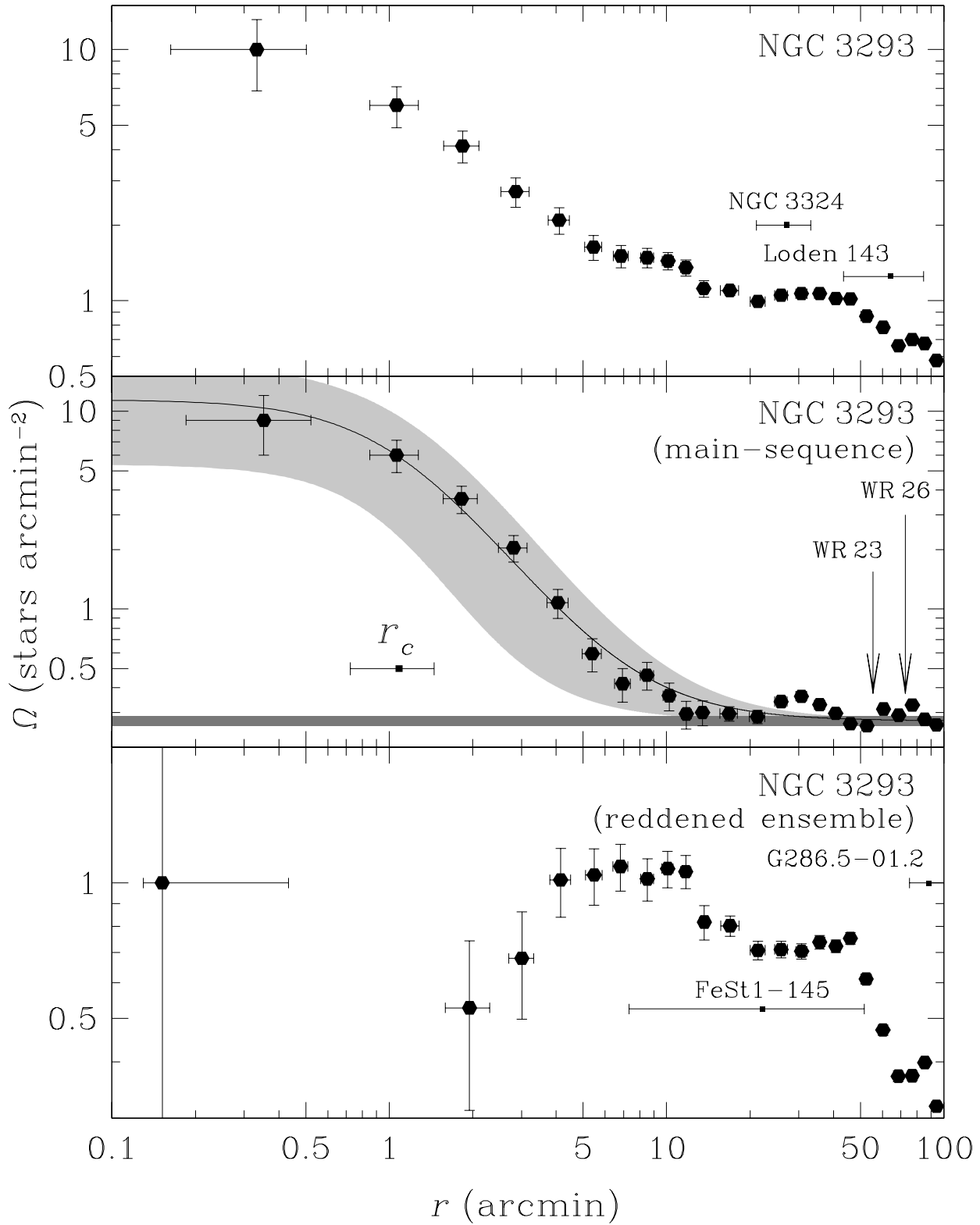}}
\caption[]{NGC\,3293: same as Fig.\,\ref{fig:cr34_rdp}.
The fitted King-like model (Eq.\,\ref{eq:king}) is shown as a solid line in the middle
panel. The core radii ($r_c$) are indicated and the 3$\sigma$ background stellar levels
(Table\,\ref{tab:rdp}) are represented by the narrow horizontal
stripes and 1$\sigma$ fit uncertainty, by the shaded regions
along the fits. The positions of nearby projected objects are indicated.}
\label{fig:ngc3293_rdp}
\end{minipage}
\end{figure*}

\begin{figure*}
\begin{minipage}{0.685\textheight}
\resizebox{\hsize}{!}{\includegraphics{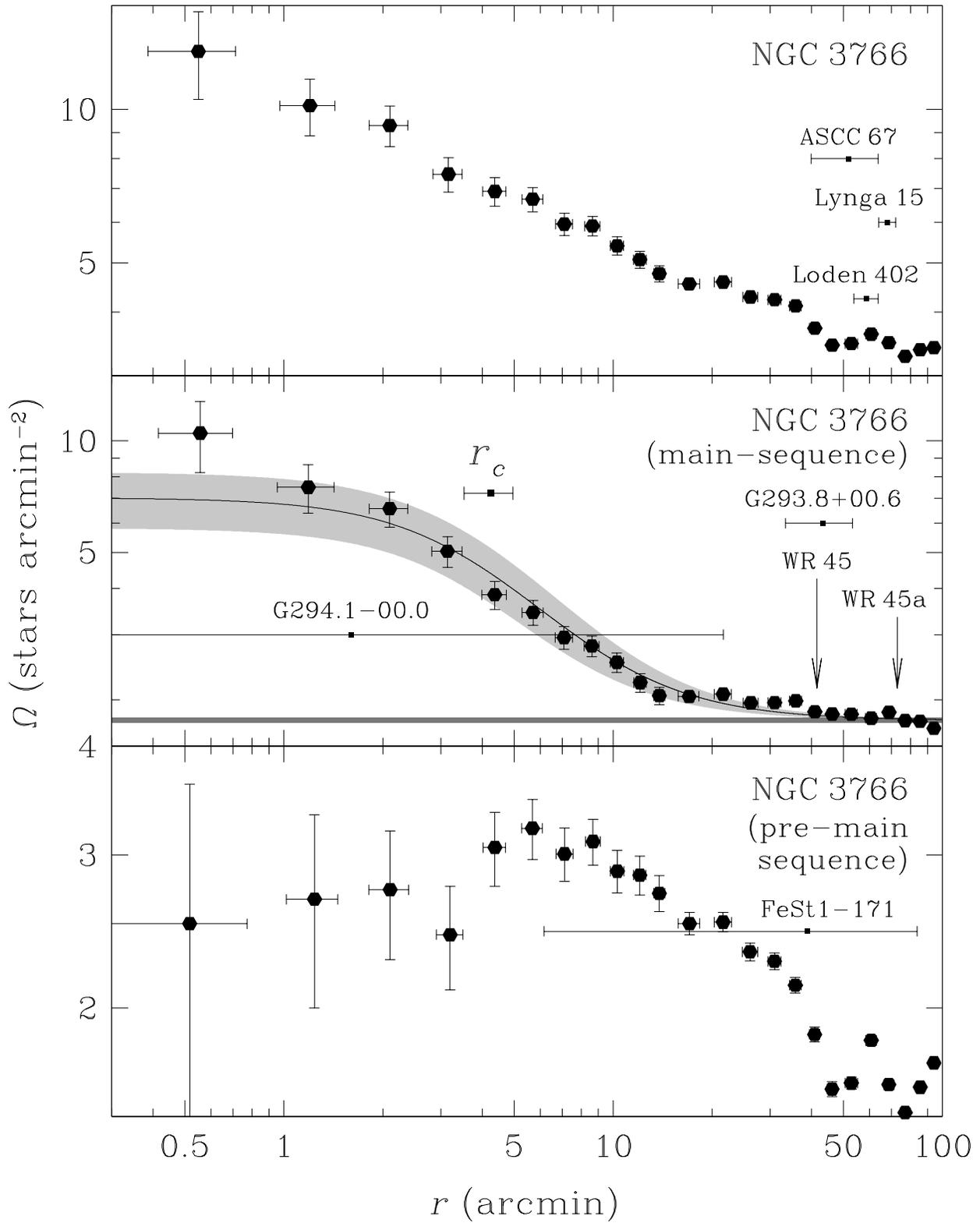}}
\caption[]{NGC\,3766: same as Fig.\,\ref{fig:ngc3293_rdp}.}
\label{fig:ngc3766_rdp}
\end{minipage}
\end{figure*}

\begin{figure*}
\begin{minipage}{0.685\textheight}
\resizebox{\hsize}{!}{\includegraphics{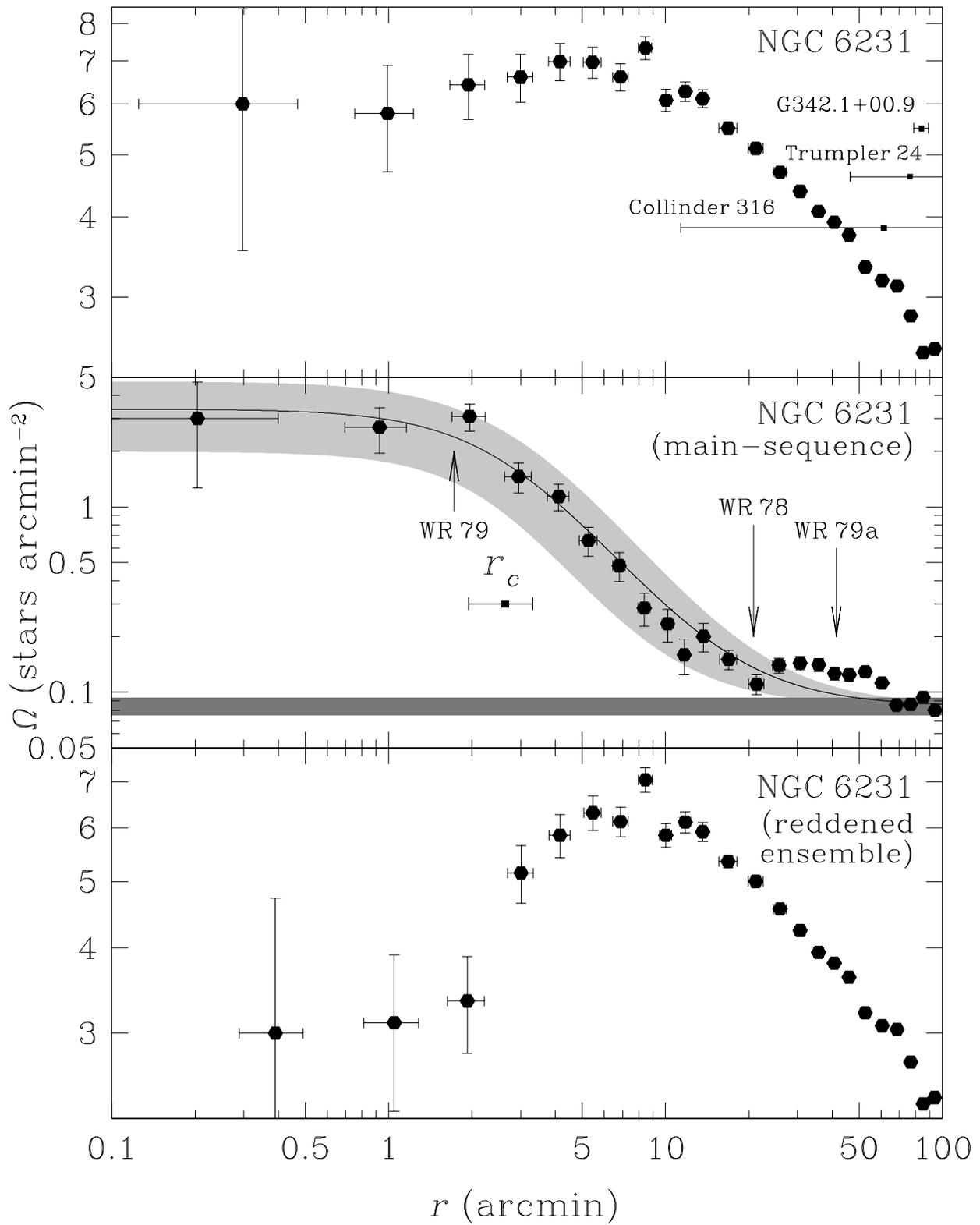}}
\caption[]{NGC\,6231: same as Fig.\,\ref{fig:ngc3293_rdp}.}
\label{fig:ngc6231_rdp}
\end{minipage}
\end{figure*}

For each cluster, we built profiles (Figs.\,\ref{fig:cr34_rdp}-\ref{fig:ngc6231_rdp})
corresponding to total (MS\,+\,PMS or probable-PMS) and partial (MS and
PMS or probable-PMS) photometry filtered by colour and magnitudes
(Sect.\,\ref{sec:cmd}). Positions of projected and surrounding objects found in catalogues
available in the Vizier database -- Wolf-Rayet stars \citep{van01}, supernova
remnants \citep{green09}, dark nebulae \citep{dubi02}, Herbig-Haro objects
\citep{reip99}, bright-rimmed clouds \citep*{sfo91} and other clusters
(\citealt{dias02}; \citealt{karc13}) -- were superimposed on
the RDPs.

We also defined a radial limit ({\it r$_{RDP}$}) for each cluster as the
projected distance from the cluster centre where the profile and the background
are indistinguishable (Table\,\ref{tab:par}). Such {\it r$_{RDP}$} may be
the intrinsic extents of the clusters \citep{bonbic07}.

Using concentric annuli and the corresponding radial distance of the most
populated position within each annulus, we built stellar RDPs for the four
clusters. Since the PMS profiles are not matched by
isothermal sphere models, we have chosen to fit only the MS profiles, except
for Collinder\,34 that has very low density of MS stars. Thus, structural
parameters (Table\,\ref{tab:rdp}) for NGC\,3293, NGC\,3766 and NGC\,6231 
were estimated by fitting their MS RDPs with a non-linear least-squares
routine to a King-like (2-parameter) model given by

\begin{equation}
{\it \Omega} (r) = {\it \Omega_{bg}} + \frac{\it \Omega_{0}}{1 + (r/r_c)^2}
\label{eq:king}
\end{equation}
where ${\it \Omega_{0}}$ is the central surface density, $r_c$ is the core radius
and ${\it \Omega_{bg}}$ is the background density measured in a surrounding
annulus (80\,$\leq$\,$r$\,$\leq$\,100\,arcmin) and kept constant.
The best fittings have a relatively small amount of scatter around the
profile, so that the root-mean-square errors are about 2 (Table\,\ref{tab:rdp}).

\begin{table*}
\begin{minipage}{1\textwidth}
\caption[]{Derived scale parameters and mass of the clusters. By columns:
(1) star cluster identification;
(2) Heliocentric distance;
(3) Galactocentric distance considering {\it R$_{\sun}$}\,=\,8.4$\pm$0.6\,kpc;
(4) angular radial limit;
(5) linear radial limit;
(6) MS stellar mass;
(7) PMS stellar mass;
(8) total stellar mass;
(9) age of the MS isochrone.}
\label{tab:par}
\renewcommand{\tabcolsep}{1.8mm}
\renewcommand{\arraystretch}{1.25}
\centering
\begin{tabular}{lcccccccc}
\hline
\hline
\multicolumn{1}{c}{cluster} &\multicolumn{1}{c}{\it d$_{\sun}$} &\multicolumn{1}{c}{\it d$_{GC}$}
&\multicolumn{1}{c}{\it r$_{RDP}$} &\multicolumn{1}{c}{\it R$_{RDP}$} &\multicolumn{1}{c}{\it M$_{MS}$}
&\multicolumn{1}{c}{\it M$_{PMS}$} &\multicolumn{1}{c}{\it M} &\multicolumn{1}{c}{\it t$_{MS}$}\\
\multicolumn{1}{c}{} &\multicolumn{1}{c}{(kpc)} &\multicolumn{1}{c}{(kpc)}
&\multicolumn{1}{c}{(arcmin)} &\multicolumn{1}{c}{(pc)} &\multicolumn{1}{c}{(M$_{\sun}$)}
&\multicolumn{1}{c}{(M$_{\sun}$)} &\multicolumn{1}{c}{(M$_{\sun}$)} &\multicolumn{1}{c}{(Myr)}\\
\multicolumn{1}{c}{(1)} &\multicolumn{1}{c}{(2)} &\multicolumn{1}{c}{(3)}
&\multicolumn{1}{c}{(4)} &\multicolumn{1}{c}{(5)} &\multicolumn{1}{c}{(6)}
&\multicolumn{1}{c}{(7)} &\multicolumn{1}{c}{(8)} &\multicolumn{1}{c}{(9)}\\
\hline
Collinder\,34 & 2.1$\pm$0.2 & 10.1$\pm$0.2 & 15.12$\pm$0.70 & 9.2$\pm$1.0 & 359$^{+37}_{-27}$ & 131 & 490$^{+37}_{-27}$ & 10\\
NGC\,3293 & 2.7$\pm$0.1 & 8.1$\pm$0.1 & 11.77$\pm$0.49 & 9.3$\pm$0.6 & 899$^{+11}_{-8}$ & - & - & 10\\
NGC\,3766 & 2.0$\pm$0.1 & 7.8$\pm$0.1 & 46.27$\pm$1.56 & 26.7$\pm$1.9 & 1682$\pm$31 & 317 & 1998$\pm$31 & 20\\
NGC\,6231 & 1.8$\pm$0.1 & 6.7$\pm$0.1 & 68.26$\pm$2.41 & 36.2$\pm$2.3 & 1939$^{+17}_{-13}$ & - & - & 10\\
\hline
\end{tabular}
\end{minipage}
\end{table*}

\begin{table*}
\begin{minipage}{1\textwidth}
\caption[]{Structural parameters obtained by fitting Eq.\,\ref{eq:king} to the MS stellar RDPs. By columns:
(1) star cluster identification;
(2) background stellar density;
(3) central stellar density;
(4) angular core radius;
(5) linear core radius;
(6) root-mean-square error.}
\label{tab:rdp}
\renewcommand{\tabcolsep}{1.4mm}
\renewcommand{\arraystretch}{1.25}
\centering
\begin{tabular}{lccccc}
\hline
\hline
\multicolumn{1}{c}{cluster} &\multicolumn{1}{c}{${\it \Omega_{bg}}$} &\multicolumn{1}{c}{${\it \Omega_0}$} &\multicolumn{1}{c}{$r_c$} &\multicolumn{1}{c}{$R_c$} &\multicolumn{1}{c}{\it E$_{rms}$}\\
\multicolumn{1}{c}{} &\multicolumn{1}{c}{(stars/arcmin$^2$)} &\multicolumn{1}{c}{(stars/arcmin$^2$)} &\multicolumn{1}{c}{(arcmin)} &\multicolumn{1}{c}{(pc)}\\
\multicolumn{1}{c}{(1)} &\multicolumn{1}{c}{(2)} &\multicolumn{1}{c}{(4)} &\multicolumn{1}{c}{(4)} &\multicolumn{1}{c}{(5)} &\multicolumn{1}{c}{(6)}\\
\multicolumn{6}{c}{\hrulefill}\\
NGC\,3293 & 0.27$\pm$0.01 & 11.18$\pm$6.06 & 1.09$\pm$0.36 & 0.9$\pm$0.3& 2.14\\
NGC\,3766 & 1.76$\pm$0.01 & 5.26$\pm$1.21 & 4.24$\pm$0.72 & 2.5$\pm$0.4 & 1.88\\
NGC\,6231 & 0.08$\pm$0.01 & 3.29$\pm$1.39 & 2.63$\pm$0.68 & 1.4$\pm$0.4 & 2.20\\
\hline
\end{tabular}
\end{minipage}
\end{table*}

\begin{figure*}
\centering
\resizebox{\hsize}{!}{\includegraphics{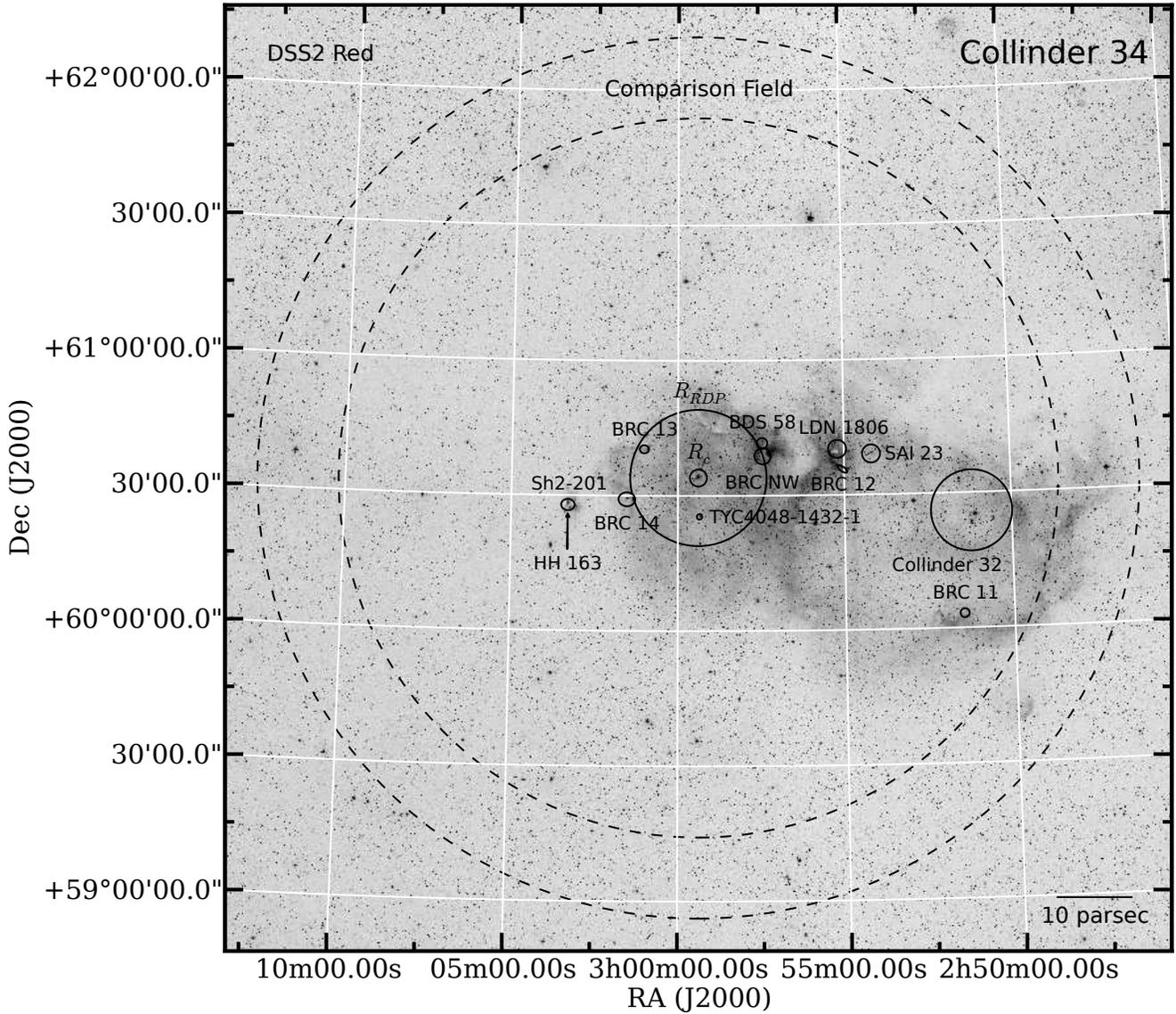}}
\caption[]{Collinder\,34: DSS2 $R$ band image including projected and neighbouring objects embedded in the IC\,1848 nebula.
$R_c$ from the total profile fitting (Table\,\ref{tab:rdp}), $R_{RDP}$ and annular field comparison area are also shown.}
\label{fig:cr34}
\end{figure*}

\begin{figure*}
\centering
\resizebox{\hsize}{!}{\includegraphics{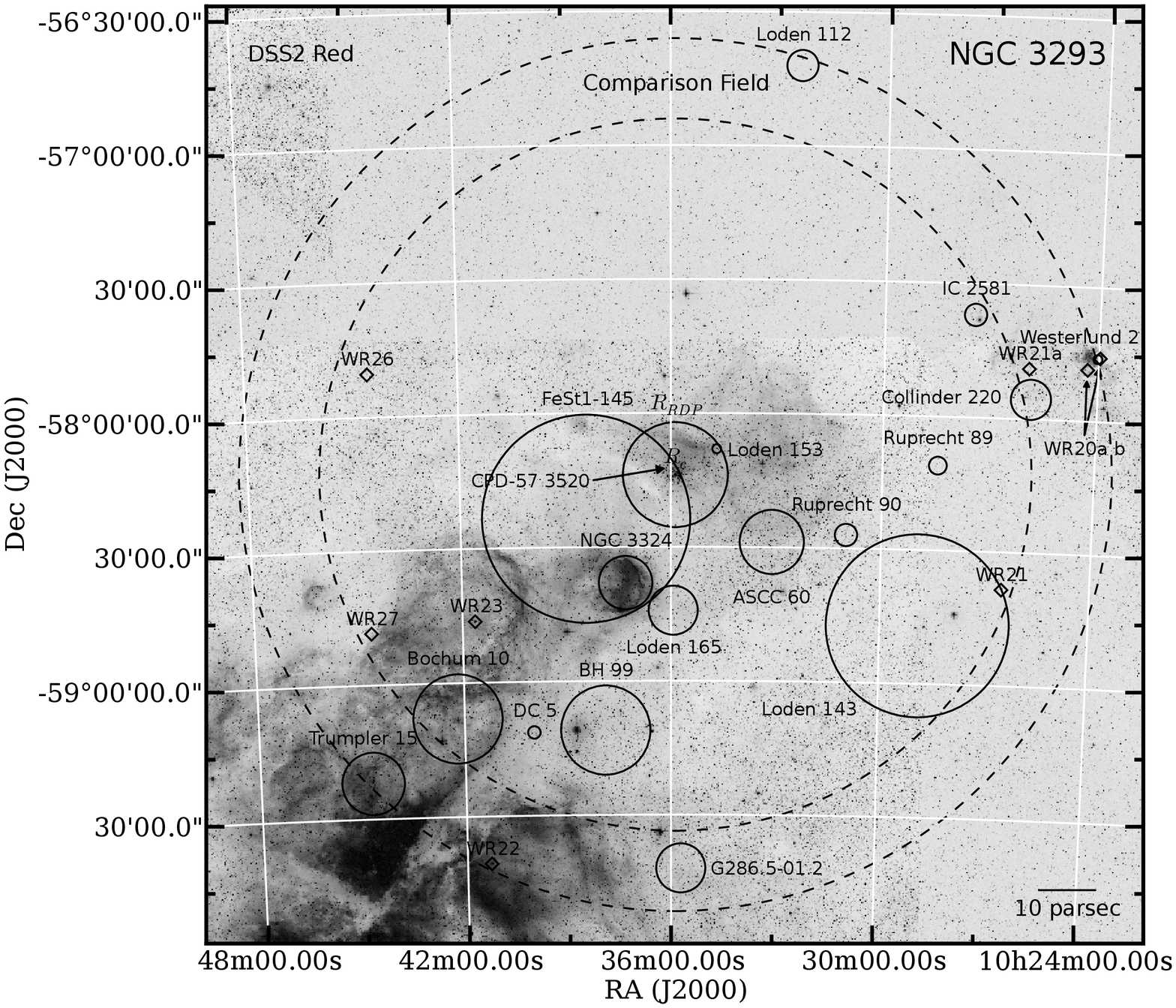}}
\caption[]{NGC\,3293: same as Fig.\,\ref{fig:cr34}.
$R_c$ from the MS profile fitting (Table\,\ref{tab:rdp}).}
\label{fig:ngc3293}
\end{figure*}

\begin{figure*}
\centering
\resizebox{\hsize}{!}{\includegraphics{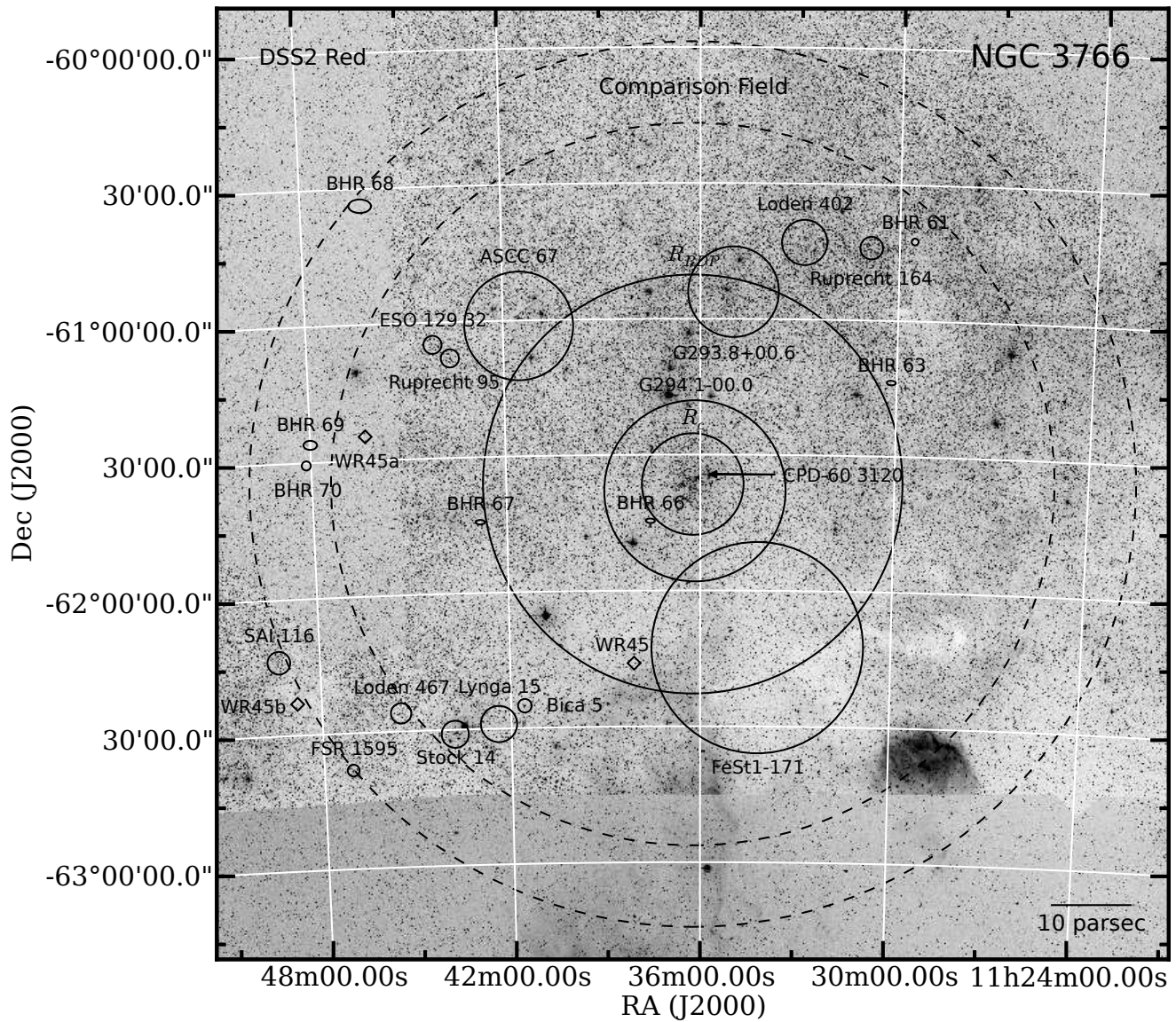}}
\caption[]{NGC\,3766: same as Fig.\,\ref{fig:cr34}.
$R_c$ from the total profile fitting (Table\,\ref{tab:rdp}).}
\label{fig:ngc3766}
\end{figure*}

\begin{figure*}
\centering
\resizebox{\hsize}{!}{\includegraphics{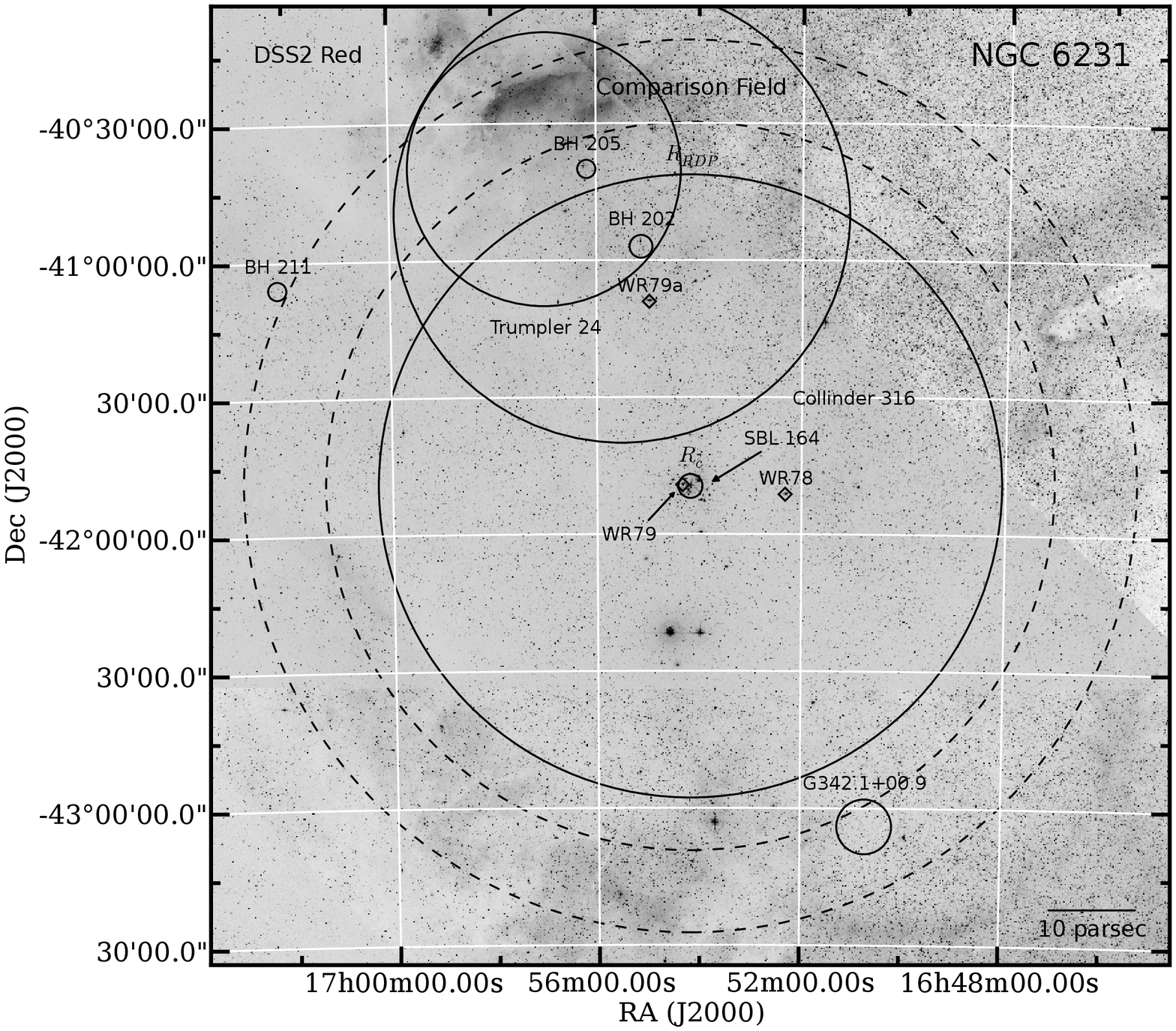}}
\caption[]{NGC\,6231: same as Fig.\,\ref{fig:cr34}.
$R_c$ from the MS profile fitting (Table\,\ref{tab:rdp}).}
\label{fig:ngc6231}
\end{figure*}

\section{Discussion}
\label{sec:dis}

A comparison of the MS structural parameters obtained in this work 
(Table\,\ref{tab:rdp}) with those available in the literature
(Table\,\ref{tab:lit}) reveals similarity within uncertainties for
NGC\,3293 and NGC\,3766. The core radius of NGC\,6231 is smaller
than that reported in the literature, but its MS profile represents
well the core and the extension structures (Fig.\,\ref{fig:ngc6231_rdp}).

\begin{table*}
\begin{minipage}{1\textwidth}
\bf
\caption[]{Cluster parameters from the literature:
\citealt{dias02} (D2002);
\citealt{pisk07} (P2007);
\citealt{buko11} (B2011);
and \citealt{karc13} (K2013).
By columns:
(1) star cluster identification;
(2) Heliocentric distance;
(3) Galactocentric distance;
(4) core radius;
(5) tidal radius;
(6) total stellar mass;
(7) age;
(8) reference.}
\label{tab:lit}
\renewcommand{\tabcolsep}{1.8mm}
\renewcommand{\arraystretch}{1.25}
\centering
\begin{tabular}{lccccccc}
\hline
\hline
\multicolumn{1}{c}{cluster} &\multicolumn{1}{c}{\it d$_{\sun}$} &\multicolumn{1}{c}{\it d$_{GC}$} &\multicolumn{1}{c}{$R_c$} &\multicolumn{1}{c}{$R_t$} &\multicolumn{1}{c}{$M$} &\multicolumn{1}{c}{Age} &\multicolumn{1}{c}{ref.}\\
\multicolumn{1}{c}{} &\multicolumn{1}{c}{(pc)} &\multicolumn{1}{c}{(pc)} &\multicolumn{1}{c}{(pc)} &\multicolumn{1}{c}{(pc)} &\multicolumn{1}{c}{(M$_{\sun}$)} &\multicolumn{1}{c}{log(Age)}\\
\multicolumn{1}{c}{(1)} &\multicolumn{1}{c}{(2)} &\multicolumn{1}{c}{(3)} &\multicolumn{1}{c}{(4)} &\multicolumn{1}{c}{(5)} &\multicolumn{1}{c}{(6)} &\multicolumn{1}{c}{(7)} &\multicolumn{1}{c}{(8)}\\
\hline
NGC\,3293 & 2327 & - & - & - & - & 7.014 & D2002\\
& 2441 & - & 1.30$\pm$0.16 & 8.69$\pm$1.04 & - & 6.75 & K2013\\
NGC\,3766 & 2218 & - & - & - & - & 7.32 & D2002\\
& 1745 & - & 1.3$\pm$0.7 &  6.9$\pm$1.4 & 2149$\pm$259 & 7.52 & P2007\\
& 1685 & - & 0.91$\pm$0.13 & 11.19$\pm$1.76 & - & 7.95 & K2013\\
NGC\,6231 & 1243 & - & - & - & - & 6.843 & D2002\\
& 1393 & - & 6.84$\pm$0.66 & 13.43$\pm$0.70 & - & 7.05 & K2013\\
\hline
\end{tabular}
\end{minipage}
\end{table*}

Notably, each cluster in this work features a detached reddened ensemble of stars in its
CMDs (Sect.\,\ref{sec:cmd}) with colour gaps $\Delta${\it (J$-$K$_S$)}
in the range 0.25-1.5\,mag (Table\,\ref{tab:dia}). In addition, the density
profiles of MS and PMS do not match each other in shape nor in extension
(Figs.\,\ref{fig:cr34_rdp}-\ref{fig:ngc6231_rdp}). We analyse whether these
particularities are connected to a second generation of PMS stars
distributed like a halo. We emphasize that the present field star
decontamination allows to access the PMS population.

Sequential star formation may be triggered by stellar winds of massive stars or
supernovae explosions. Two mechanisms were proposed to explain this process:
collect-and-collapse \citep{ella77}
and radiation-driven implosions \citep*{swk82}.
Both may operate simultaneously in a star forming region, but the former
is an important mechanism of massive star formation because the expanding
\mbox{H\,{\sc ii}} region sweeps out the cloud collecting material as a
snowball \citep{zav06}.

Signposts of collect-and-collapse and radiation-driven implosions remain for
a while. Cometary globules such as CG7S \citep{leflaz95}, and
bright-rimmed clouds such as BRC\,14 \citep{lorwoo78}, both in the
complex of the Heart and Soul nebulae (that embeds Collinder\,34),
may be characteristic of radiation-driven implosions and collect-and-collapse,
respectively.

Such substructures, further neighbouring and projected objects in the
area of the sample clusters are shown in Figs.\,\ref{fig:cr34}-\ref{fig:ngc6231}
with Digitized Sky Survey 2 (DSS2) $R$ band images generated by {\it SkyView} \citep{sk96}.

In the following, we carry out individual analyses of each cluster.

\subsection{Collinder\,34}
\label{sec:cr34}

Collinder\,34 (Fig.\,\ref{fig:cr34}) is part of the complex of
the Heart (Sh2-190) and Soul (Sh2-199) nebulae,
a star forming region ionized by the Cas\,OB6 association.

Originally catalogued by \citet{col31} at the B1900 coordinates
$\alpha$\,=\,02$^{\rmn{h}}$43$^{\rmn{m}}$30$^{\rmn{s}}$
and $\delta$\,=\,+60\degr01\arcmin00\arcsec,
it was recently identified as SAI\,24 \citep{glu10}. Actually,
Collinder\,34, together with Collinder\,32 (also known as the
IC\,1848 cluster), is inserted in a hierarchical structure of
the Collinder\,33 association.

Collinder\,34 is located inside a bubble in the ''baby's head'' of the
Sh2-199 nebula, centred in the O-type star HD\,18326. It is
surrounded by some infrared clusters -- embedded in Sh2-201, BRC\,13 and
BRC\,14 \citep{chau11} -- together with the small optically visible cluster
BDS\,58 \citep{bdsb03}.

Among the clusters in the present study, Collinder\,34 has the narrowest colour gap
between MS and PMS (Fig.\,\ref{fig:cr34_cmd}),
{\it $\Delta$(J$-$K$_S$)}\,$\sim$\,0.25\,mag.
Its PMS population shows some dispersion in the 2CD
(Fig.\,\ref{fig:cr34_2cd}) and a small near-IR excess,
typically associated with optically thin dust circumstellar disks
featured by
Weak-lined T Tauri stars (WTTSs). This is consistent with the location of
the bulk of the PMS in a region limited by {\it (J$-$H)$_0$}\,$\lesssim$\,0.5
and {\it (H$-$K$_S$)$_0$}\,$\lesssim$\,0.5 \citep{meye97}. In addition,
a few points are located in {\it (H$-$K$_S$)$_0$}\,$>$\,0.5, a region
usually occupied by Herbig Ae/Be stars \citep{her05}.

Stellar RDPs of Collinder\,34 (Fig.\,\ref{fig:cr34_rdp}) show a structure
dominated by the PMS with its MS stars concentrated in the core. The MS
profile presents an excess near the centre that cannot be
interpreted as core collapse in such a young cluster. This may be
characteristic of primordial mass segregation caused by competitive accretion
that preferentially forms massive stars in the deepest part of the
cluster potential well \citep{bon01}.

\subsection{NGC\,3293}
\label{sec:ngc3293}

Projected close to the $\eta$\,Carinae nebula, the star cluster NGC\,3293
(Fig.\,\ref{fig:ngc3293}) forms an interacting pair with NGC\,3324
(\citealt{fuente09}; \citealt{fuente10}).

The sequences of NGC\,3293 are separated in the decontaminated CMDs
(Fig.\,\ref{fig:ngc3293_cmd}) by a gap of
{\it $\Delta$(J$-$K$_S$)}\,$\sim$\,1.5\,mag, with a heavily reddened ensemble
of stars. While, its RDPs (Fig.\,\ref{fig:ngc3293_rdp}) show
a centrally concentrated structure of MS stars well
fitted by Eq.\,\ref{eq:king} (Table\,\ref{tab:rdp}) with the reddened
ensemble surrounding the core, like a halo.

An inspection of the 2CD (Fig.\,\ref{fig:ngc3293_2cd}) suggests that
a fraction of these reddened stars might be background dwarfs and giants
that were not removed by the decontamination procedure (Sect.\,\ref{sec:dec}).
Apparently, any nearby area was not adequate to model the field in detail.
Anyway, the bulk of the reddened ensemble is outside the strip of dwarfs and giants,
and might be formed by T Tauri stars with an absorption of $A_V$\,$\sim$\,3.5\,mag.

\subsection{NGC\,3766}
\label{sec:ngc3766}

The cluster NGC\,3766 (Fig.\,\ref{fig:ngc3766}) in the Carina complex
includes the supernova remnant G294.1-00.0 \citep{green09}
and at least 16 classical Be stars \citep*{mcsw09}.

Decontaminated CMDs of NGC\,3766 (Fig.\,\ref{fig:ngc3766_cmd}) show
a colour gap of {\it $\Delta$(J$-$K$_S$)}\,$\sim$\,0.5\,mag separating a
broad MS and an elongated PMS. The shock wave of the supernova
G294.1-00.0 might have removed the circumstellar disc of
part of the PMS stars causing that elongation. In fact,
an examination of the 2CD (Fig.\,\ref{fig:ngc3766_2cd}) reveals a PMS
with little or no near-IR excess, consistent with discless T Tauri stars.

Regarding the cluster structure, the stellar RDPs of NGC\,3766 (Fig.\,\ref{fig:ngc3766_rdp})
have a well fitted King-like profile of MS stars (Table\,\ref{tab:rdp}) in a centrally
concentrated structure with excess near the centre, similar to Collinder\,34.
In addition, its PMS stars surround the MS core, suggesting a halo.

\subsection{NGC\,6231}
\label{sec:ngc6231}

The cluster NGC\,6231 (Fig.\,\ref{fig:ngc6231}),
at the core of the Sco\,OB1 association \citep{per90}, is rich in
early-type stars (at least 150) and
more than half of its OB stars are members of binary systems (\citealt{rab96};
\citealt{sana08}).

NGC\,6231 is inside a large bubble in the Gum\,55 \mbox{H\,{\sc ii}} region.
Stellar winds of massive stars, like WR\,79 \citep{van01},
or past supernovae explosions might have depleted the central area
of the cloud resulting in a low differential reddening and
a deficit of H$\alpha$ emission in the PMS stars \citep*{sun98}.

Decontaminated CMDs of NGC\,6231 (Fig.\,\ref{fig:ngc6231_cmd}) show a
narrow MS separated from a heavily reddened ensemble of stars by a wide colour
gap of {\it $\Delta$(J$-$K$_S$)}\,$\sim$\,1.0\,mag,
like that of NGC\,3293 (Sect.\,\ref{sec:ngc3293}).

The 2CD of NGC\,6231 (Fig.\,\ref{fig:ngc6231_2cd}) shows that the reddened ensemble
is placed along a broad band with a high absorption of $A_V$\,$\sim$\,3.5\,mag
for the bulk. Furthermore, similar to NGC\,3293 (Sect.\,\ref{sec:ngc3293}),
background dwarfs and giants seem to have survived to decontamination.

Finally, the stellar MS RDP of NGC\,6231 (Fig.\,\ref{fig:ngc6231_rdp}) is
well fitted by Eq.\,\ref{eq:king} (Table\,\ref{tab:rdp}). Such profile matches
that by \citet*{sun13} which extents to 20\,arcmin for high mass stars and
shows an extension in 20\,$<$\,$r$\,$<$\,60\,arcmin. 

\subsection{Diagnostic diagrams}
\label{sec:comp}

\begin{figure}
\begin{minipage}{0.435\textwidth}
\resizebox{\hsize}{!}{\includegraphics{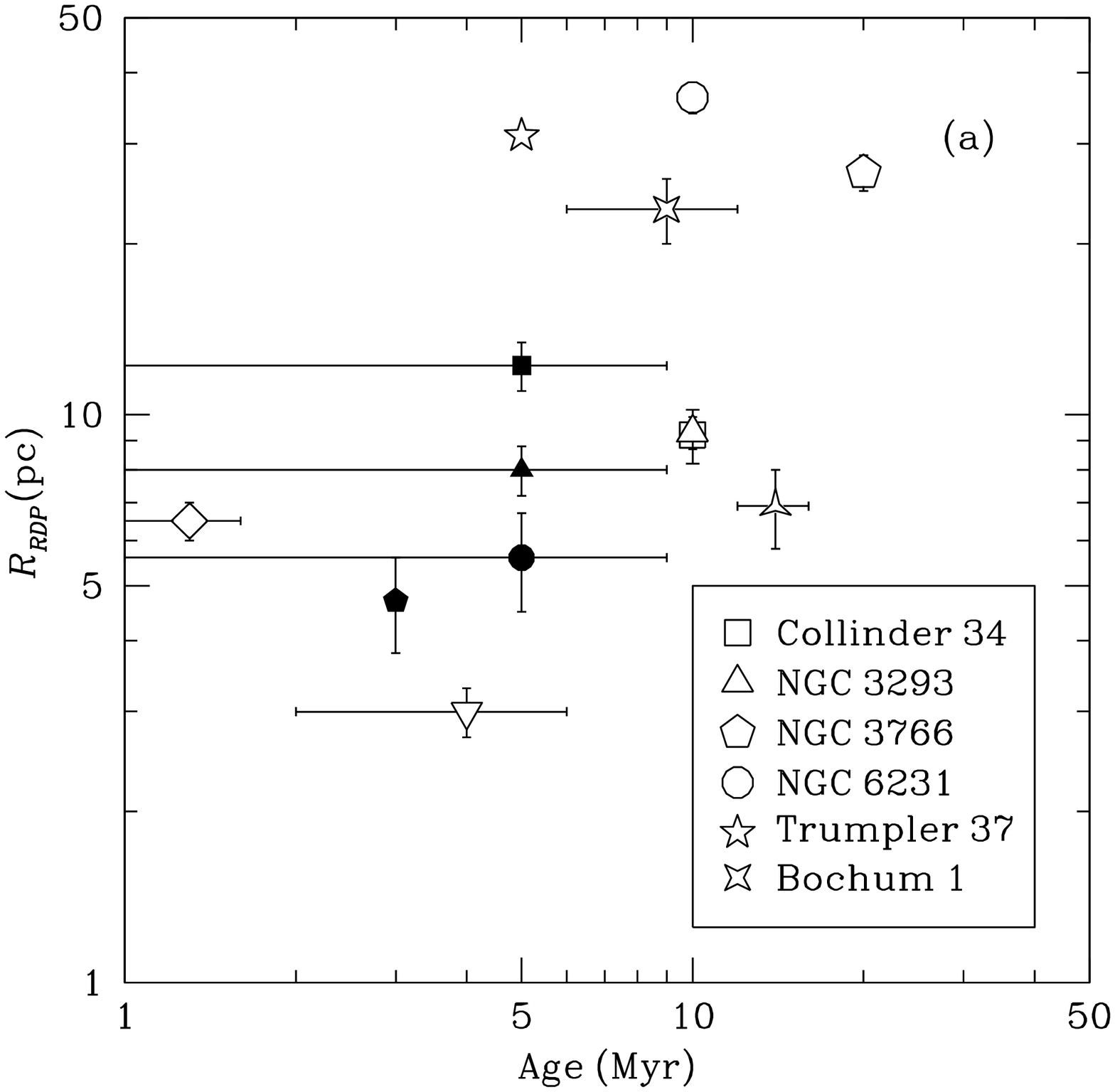}}
\resizebox{\hsize}{!}{\includegraphics{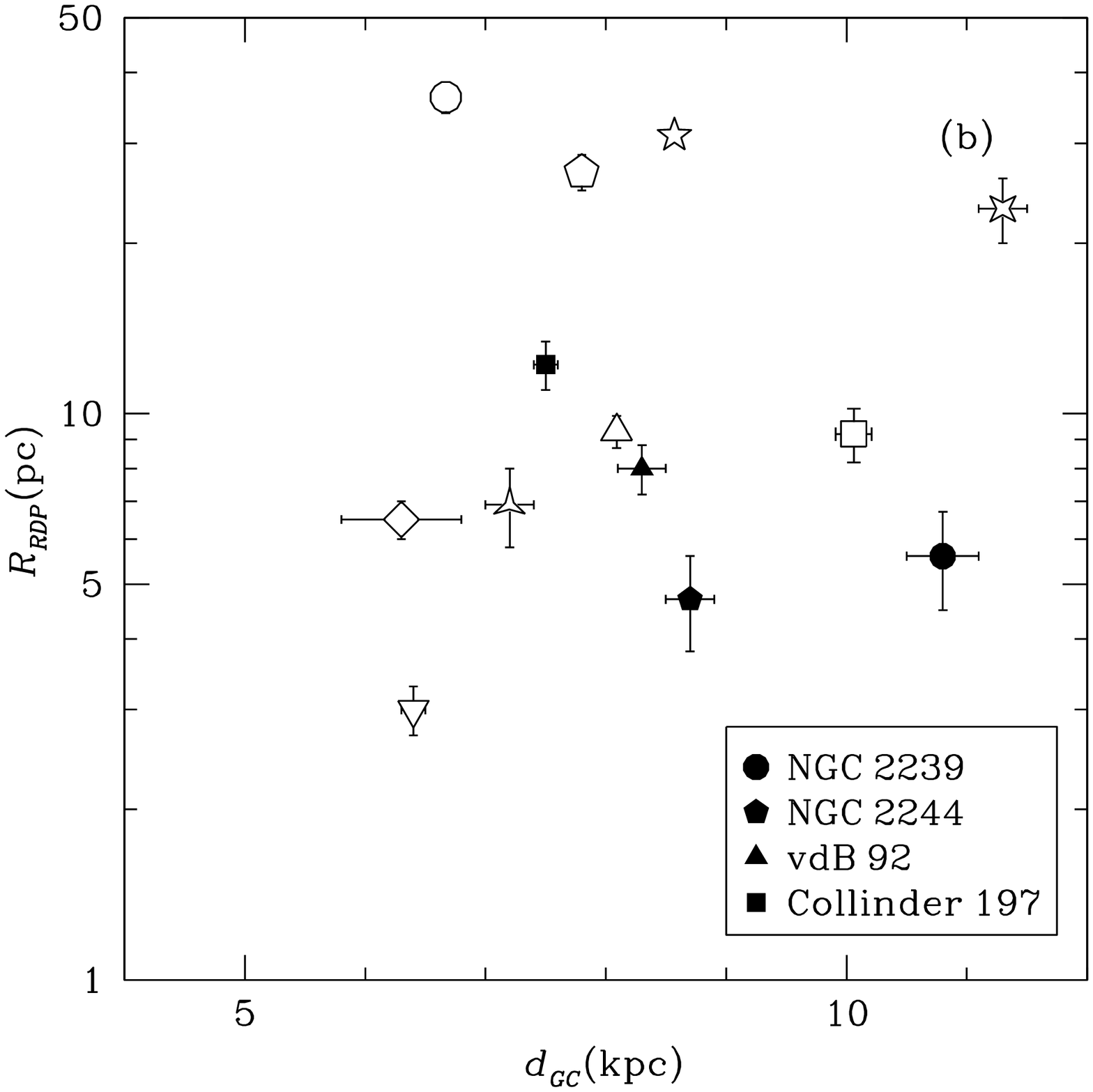}}
\resizebox{\hsize}{!}{\includegraphics{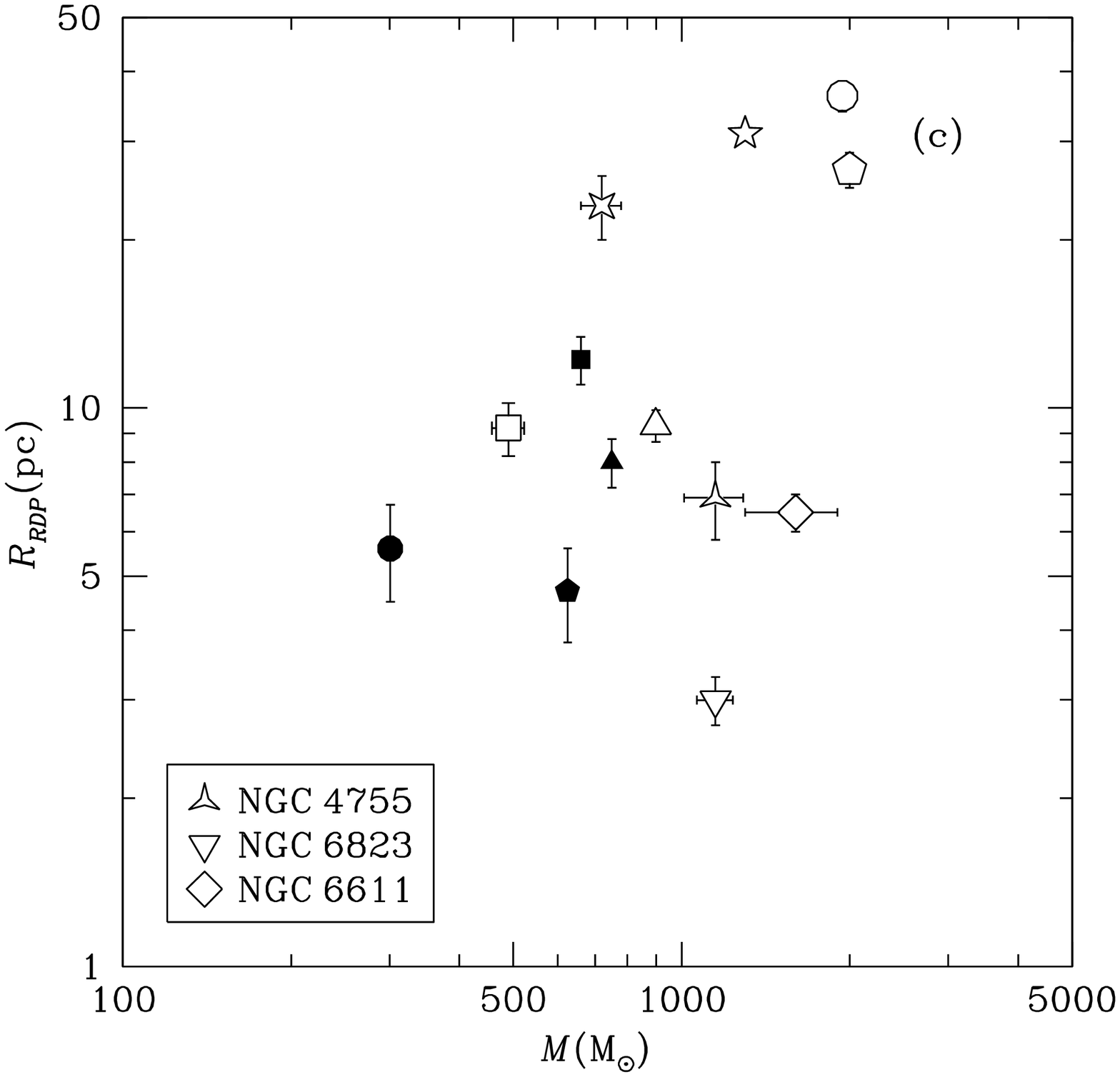}}
\caption[]{Diagnostic diagrams comparing cluster parameters. Some error bars are smaller than the symbols.}
\label{fig:diag}
\end{minipage}
\end{figure}

In order to compare the derived cluster parameters -- radial limit, age, Galactocentric
distance and mass -- of the present sample with those of other clusters, we
built diagnostic diagrams (Fig.\,\ref{fig:diag}) for a selected sample
of young objects within the age range 1\,Myr to 14\,Myr. They are:
NGC\,4755 \citep{bbob06};
NGC\,6611 \citep*{bsb06};
Bochum\,1 and NGC\,6823 \citep*{bbd08};
NGC\,2244 and NGC\,2239 \citep{bonbic09};
Collinder\,197 and vdB\,92 \citep{bonbic10};
and Trumpler\,37 \citep{sbb12}.

The radius time evolution of the clusters seems to be constrained by the
Galactocentric distance and mass, as seen in Fig.\,\ref{fig:diag}, panels (a),
(b) and (c). The more massive clusters NGC\,3766 and NGC\,6231 appear to keep
most of their stars despite the relatively smaller Galactocentric distance.
They have comparable radii to those of Bochum\,1 and
Trumpler\,37, systems that seem to be evolving to OB associations. On the other
hand, Collinder\,34 and NGC\,3293 occupy intermediate positions in the diagrams,
mixed to the ordinary embedded clusters.
Our sample is inserted into a scale of early evolving star clusters
tending to dissolution, as expected for most of the embedded clusters
\citep{lada03}.

\begin{table*}
\begin{minipage}{1\textwidth}
\caption[]{Star cluster RDPs diagnostic. By columns:
(1) star cluster identification;
(2) MS RDP morphology;
(3) PMS RDP morphology;
(4) extended RDP;
(5) RDP with central excess;
(6) {\it (J$-$K$_S$)} gap between MS and PMS.}
\label{tab:dia}
\renewcommand{\tabcolsep}{1.8mm}
\renewcommand{\arraystretch}{1.25}
\centering
\begin{tabular}{lccccc}
\hline
\hline
\multicolumn{1}{c}{cluster} &\multicolumn{1}{c}{RDP$_{MS}$} &\multicolumn{1}{c}{RDP$_{PMS}$}
&\multicolumn{1}{c}{RDP$_{ext}$} &\multicolumn{1}{c}{RDP$_{exc}$} &\multicolumn{1}{c}{\it $\Delta$(J$-$K$_S$)}\\
\multicolumn{1}{c}{(1)} &\multicolumn{1}{c}{(2)} &\multicolumn{1}{c}{(3)}
&\multicolumn{1}{c}{(4)} &\multicolumn{1}{c}{(5)} &\multicolumn{1}{c}{(6)}\\
\hline
Collinder\,34 & non-King & King-like & no & yes & 0.25 \\
NGC\,3293 & King-like & non-King & yes & no & 1.5 \\
NGC\,3766 & King-like & non-King & yes & yes & 0.5 \\
NGC\,6231 & King-like & non-King & yes & no & 1.0 \\
\hline
\end{tabular}
\end{minipage}
\end{table*}

\section{Concluding remarks}
\label{sec:con}

We have studied four young star clusters and found CMDs with
detached PMS from the MS at least for two objects. Colour gaps between these
evolutionary sequences were revealed by field star decontamination
-- 0.25\,mag for Collinder\,34, and 0.5\,mag for NGC\,3766. This
might be evidence of sequential star formation.

On the other hand, the clusters NGC\,3293 and NGC\,6231 have wider colour
gaps, 1.5 and 1.0\,mag, respectively. An inspection of their 2CDs suggests
that those heavily reddened ensembles appear to be contaminated by residual background
dwarfs and giants of the field. Anyway, the bulk of those ensembles might
consist of very reddened PMS stars.

Spatial structures of the sequences of each cluster were examined by
means of stellar RDPs revealing that the detached sequences are distributed in
different ways for each cluster. The MS profiles of NGC\,3293, NGC\,3766 and
NGC\,6231 have been fitted by a King-like model yielding 
similar structural parameters as compared to the literature.
Furthermore, the RDPs of Collinder\,34 and NGC\,3766 present central
excesses that may represent primordial mass segregation.

Finally, this study suggests that reddened halos of PMS stars as formed
from the parental cloud may still be present after 5-20\,Myr.

\section*{Acknowledgments}
We acknowledge support from the Brazilian Institution CNPq.
We thank an anonymous referee for constructive comments and suggestions.
This research has made use of the WEBDA database, operated at the Department of Theoretical Physics and Astrophysics of the Masaryk University.
We acknowledge the use of NASA's {\it SkyView} facility (http://skyview.gsfc.nasa.gov) located at NASA Goddard Space Flight Center.
This research has made use of the VizieR catalogue access tool, CDS, Strasbourg, France.
This research has made use of the SIMBAD database, operated at CDS, Strasbourg, France.
This publication makes use of data products from the Two Micron All Sky Survey, which is a joint project of the University of Massachusetts and the Infrared Processing and Analysis Center/California Institute of Technology, funded by the National Aeronautics and Space Administration and the National Science Foundation.

\label{lastpage}

\clearpage

\end{document}